\documentclass[11pt,letterpaper]{article}
\usepackage{opex3}

\usepackage{graphicx}
\usepackage{dcolumn}
\usepackage{bm}
\usepackage{subfig}
\usepackage{t1enc}
\usepackage{amsmath}
\usepackage[noadjust]{cite}

\newcommand{\figref}[1]{\text{Fig.~\ref{#1}}}
\newcommand{\secref}[1]{\text{Section~\ref{#1}}}

\begin{document}

\title{Graphene-based plasmonic switches at near infrared frequencies}

\author{J.~S.~G\'omez-D\'{\i}az and J.~Perruisseau-Carrier}

\address{Adaptive MicroNanoWave Systems, LEMA/Nanolab \\ École Polytechnique Fédérale de Lausanne, 1015 Lausanne, Switzerland}

\email{juan-sebastian.gomez@epfl.ch, julien.perruisseau-carrier@epfl.ch} 



\begin{abstract}
The concept, analysis, and design of series switches for graphene-strip plasmonic waveguides at near infrared frequencies are presented. Switching is achieved by using graphene's field effect to selectively enable or forbid propagation on a section of the graphene strip waveguide, thereby allowing good transmission or high isolation, respectively. The electromagnetic modeling of the proposed structure is performed using full-wave simulations and a transmission line model combined with a matrix-transfer approach, which takes into account the characteristics of the plasmons supported by the different graphene-strip waveguide sections of the device. The performance of the switch is evaluated versus different parameters of the structure, including surrounding dielectric media, electrostatic gating and waveguide dimensions.
\end{abstract}



\section{Introduction}

The field of plasmonics represents a new exciting area for the control of electromagnetic waves at scales much smaller than the wavelength. It is based on the propagation of surface plasmon polaritons (SPPs) \cite{Pendry2004, Pitarke07, Wang11}, which are electromagnetic waves propagating along the surface interface between a metal (or a semiconductor) and a dielectric. SPPs are typically obtained in the visible range by using noble metal such as gold or silver \cite{Pitarke07}, but they are also supported at lower frequencies by composite materials \cite{Pendry2004, Elser07}. Surface plasmons have served as a basis for the development of nanophotonic devices \cite{Barnes03}, merging the fields of photonics and electronics at the nanoscale \cite{Pitarke07} and finding application in different areas such as imaging \cite{Kawata09} or sensing \cite{Homola99}.

Graphene \cite{Geim2007} provides excellent possibilities to dynamically manipulate electromagnetic waves \cite{Crassee10, Engheta11, Tamagnone12_apl}. Its unique electric properties, which can be controlled by simply applying an external magnetostatic or electrostatic field, allows the propagation of surface plasmons in terahertz and infra-red frequency bands \cite{Grigorenko12}. Compared to conventional materials, such as silver or gold, SPPs on graphene present important advantages \cite{Koppens11} including tunability, low-losses, and extreme mode confinement. Several authors have studied the characteristics of plasmons propagating along $2$D graphene sheets \cite{Hanson08, Jablan09, Ferreira12, Sebas12_jap} and ribbons/strips \cite{Nikitin11, Sounas11, Christensen12}, and different configurations have already been proposed to enhance their guiding properties \cite{Hwang07}.

The ability to allow or to forbid the propagation of SPPs on these structures is a key building block for future plasmonic-based devices. Graphene-based switches have already been proposed in the literature at DC \cite{Standley08, Palacios10} and microwave frequencies \cite{Milaninia09, Perruisseau12_LAPC}, based either on graphene electric field effect or exploiting the electromechanical properties of graphene. In the optical regime, \cite{Bludov10} proposed a structure able to switch the reflectance of a plane wave incoming from free-space between two different states, namely total reflection and total absorption. This is obviously a different functionality from the switching of a guided plasmonic wave as concerned here. Finally, in a recent work \cite{Chen12_phase_shifters}, graphene-based longitudinally homogeneous parallel-plate waveguides were proposed to obtain switches and phase-shifter in the low terahertz band.

In this context, we propose and design series switches able to control the propagation of surface plasmons on finite graphene strips at near infrared frequencies. The structures are composed of a chemically-doped graphene strip, host waveguide of the switch, transferred onto a dielectric and of three polysilicon gating pads beneath the strip. The switch consists of the central section of the host waveguide, whereas the outer sections connect the switch to the input and output ports of the device. Switching is achieved by modifying the gate voltage of the central pad, which in turns controls the guiding properties of the strip in the area of the switch. In the ON state, the whole host waveguide has the same propagating characteristics and the structure behaves as a simple plasmonic transmission line (TL) propagating the incoming energy towards the output port. In the OFF state, the guiding properties of the central waveguide section are modified to provide large isolation between input and output ports. The structures are characterized by applying a transmission line approach and by using the commercial full-wave software HFSS. Note that \cite{Chen12_phase_shifters} recently studied graphene-based longitudinally inhomogeneous parallel-plate waveguide using solely TL techniques. Here, we present a rigorous comparison between the TL approach and a full-wave solver for the case of finite graphene-strip waveguides, demonstrating that while the former provides extremely fast results and physical insight into the structure, its mono-modal nature may lead to inaccuracies when characterizing the OFF state of the device. Finally, several devices, based on ideal $2$D graphene surfaces and on realistic finite strips, are discussed and studied, evaluating their performance versus different parameters of the switches.

\section{Implementation and modeling}
This section details the concept, implementation and modeling of the proposed graphene-based switches. We first briefly review the characteristics of surface plasmons polaritons propagating on ideal 2D graphene surfaces \cite{Jablan09} and on finite strips \cite{Nikitin11,Christensen12}. Then, we describe the proposed switches, detailing their underlying operating principle and discussing its potential technological implementation. Finally, we address the modeling of the different structures, using both a TL approach and full-wave commercial software.

\subsection{Characteristics of TM surface plasmon-polaritons}
Graphene is a one atom thick gapless semiconductor \cite{Geim2007, Novoselov04}, which can be characterized by a complex surface conductivity $\sigma$. This conductivity can be modelled using the well-known Kubo formalism \cite{Gusynin09}, and mainly depends on graphene chemical potential $\mu_c$, which may be controlled by modifying the initial doping of the material or by applying an external electrostatic field, and on the phenomenological scattering rate $\Gamma$:
\begin{align}
\sigma(\omega,\mu_c,\Gamma,T)=\frac{j q_e^2(\omega-j2\Gamma)}{\pi \hbar^2}&\left[\frac{1}{(\omega-j2\Gamma)^2}\int_0^{\infty} \epsilon \left(\frac{\partial f_d(\epsilon)}{\partial \epsilon}-\frac{\partial f_d(-\epsilon)}{\partial \epsilon}\right)\partial\epsilon \right. \nonumber \\ &- \left. \int_0^{\infty} \frac{f_d(-\epsilon)-f_d(\epsilon)}{(\omega-j2\Gamma)^2-4(\epsilon/\hbar)^2}\partial\epsilon\right], \label{eq: Conductivity}
\end{align}
where $\omega$ is the radial frequency, $\epsilon$ is energy, $T$ is temperature, $-q_e$ is the charge of an electron, $\hbar$ is reduced Planck's constant, $k_B$ is Boltzmann's constant, and $f_d$ is the Fermi-Dirac distribution:
\begin{equation}
f_d(\epsilon)=\left(e^{(\epsilon-|\mu_c|)/k_BT}+1\right)^{-1}.
\label{eq_Fermi_distribution}
\end{equation}
Note that the first and second terms of Eq.~(\ref{eq: Conductivity}) are related to intraband and interband contributions of graphene conductivity, respectively \cite{Gusynin09}. The real part of conductivity is sensitive to $\Gamma$ at the frequencies where intraband contributions of graphene dominate (i.e. $\hbar\omega/\mu_c\ll1$) while it mainly depends on temperate in the frequency region where interband contributions dominate (i.e. $\hbar\omega/\mu_c\geq1$). In addition, note that the conductivity model employed here does not depend on the wavevector, i.e. spatial-dispersion effects are neglected \cite{Falkovsky07b, sebas12_Spatial_dispersion}. However, in the frequency range considered  ($25-30$~THz, where interband contributions of conductivity are non-negligible) and for the parameters employed in this work, the influence of this phenomena on graphene conductivity is small \cite{Jablan09}, and the approximate model of graphene conductivity of Eq.~(\ref{eq: Conductivity}) leads to accurate results.

One of the main advantages of graphene is that its chemical potential can be tuned over a wide range (typically from $-1$~eV to $1$~eV) by applying a transverse electric field via a DC biased gated structure, thus allowing to control graphene conductivity. The applied DC voltage ($V_{DC}$) modifies graphene carrier density ($n_s$) as
\begin{equation}
C_{ox}V_{DC}=q_en_s,
\label{Capacitance_and_carrier_densitiy_relation}
\end{equation}
where $C_{ox}=\varepsilon_r\varepsilon_0/t$ is the gate capacitance, and $\varepsilon_r$ and $t$ are the permittivity constant and thickness of the gate dielectric. Note that Eq.~(\ref{Capacitance_and_carrier_densitiy_relation}) neglects graphene quantum capacitance \cite{Chen08}, which may be important in case of extremely thin dielectrics ($t\sim$nms) or very high dielectric constants. In addition, the carrier density is related with $\mu_c$ via the expression
\begin{equation}
n_s=\frac{2}{\pi\hbar^2v_f^2}\int_0^\infty\epsilon[f_d(\epsilon-\mu_c)-f_d(\epsilon+\mu_c)]\partial\epsilon,
\label{eq:Carrier_densitiy}
\end{equation}
where $v_f$ is the Fermi velocity ($\sim 10^8$ cm/s in graphene). Then, the desired chemical potential $\mu_c$ is accurately retrieved by numerically solving Eqs.~(\ref{Capacitance_and_carrier_densitiy_relation}) and (\ref{eq:Carrier_densitiy}). An approximate closed-form expression to relate $\mu_c$ and $V_{DC}$ is given by
\begin{equation}
\mu_c\approx\hbar v_f\sqrt{\frac{\pi C_{ox} V_{DC}}{q}},
\label{eq:mu_c_vs_V_DC}
\end{equation}
which is obtained by combining Eq.~(\ref{Capacitance_and_carrier_densitiy_relation}) with the graphene carrier density at zero temperature \cite{Berdebes09}
\begin{equation}
n_s=\frac{1}{\pi}\left(\frac{\mu_c}{\hbar v_f}\right)^2.
\label{eq:Carrier_densitiy_approx}
\end{equation}

The characteristics of SPPs supported by graphene depend on the conductivity of the material and on the type of waveguide employed. In the case of ideal $2$D graphene sheets, the dispersion relation of the propagating modes can be obtained as \cite{Jablan09}
\begin{equation}
\frac{\omega\varepsilon_{r_1}\varepsilon_{0}}{\sqrt{\varepsilon_{r_1}k_0^2-k_\rho^2}}-\frac{\omega\varepsilon_{r_2}\varepsilon_{0}}{\sqrt{\varepsilon_{r_2}k_0^2-k_\rho^2}}=-\sigma,
\label{eq:Dispersion_TM_general}
\end{equation}
where $\varepsilon_{0}$ is the vacuum permittivity, $\varepsilon_{r_1}$ and $\varepsilon_{r_2}$ are the dielectric permittivities of the media surrounding graphene, $k_0=\omega/c$ is the free space wavenumber and $k_\rho=\beta-j\alpha$ is the complex propagation constant of the SPP mode, being $\beta$ and $\alpha$ the phase and atenuation constants, respectively. In addition, the characteristic impedance of the SPP may be expressed as \cite{Pozar05}
\begin{equation}
Z_C=\frac{k_\rho}{\omega\varepsilon_0\varepsilon_{\text{eff}}},
\label{eq:Characteristic_impedance}
\end{equation}
where $\varepsilon_{\text{eff}}$ is the effective permitivitty constant of the surrounding medium. Note that though the characteristic impedance is not often employed to model SPPs \cite{Pitarke07}, this parameter will be useful to understand and optimize the behavior of the proposed switches.

In the case of finite graphene strips, the dispersion relation of propagating surface plasmons cannot be derived analytically and one has to resort to purely numerically full-wave solvers. There are two different types of SPP propagating along the strip \cite{Nikitin11, Christensen12}: the waveguide type, which has the field concentrated along the whole strip, and the edge type, where the field is focused on the rims of the strip. Note that graphene relaxation time mainly control the propagation length of the modes, barely affecting to their field confinement. In addition, the characteristics of these modes can also be tuned by modifying the chemical potential of graphene.
\begin{figure} \centering
\subfloat[]{\label{fig:_Re_kp}
\includegraphics[width=0.45\columnwidth]{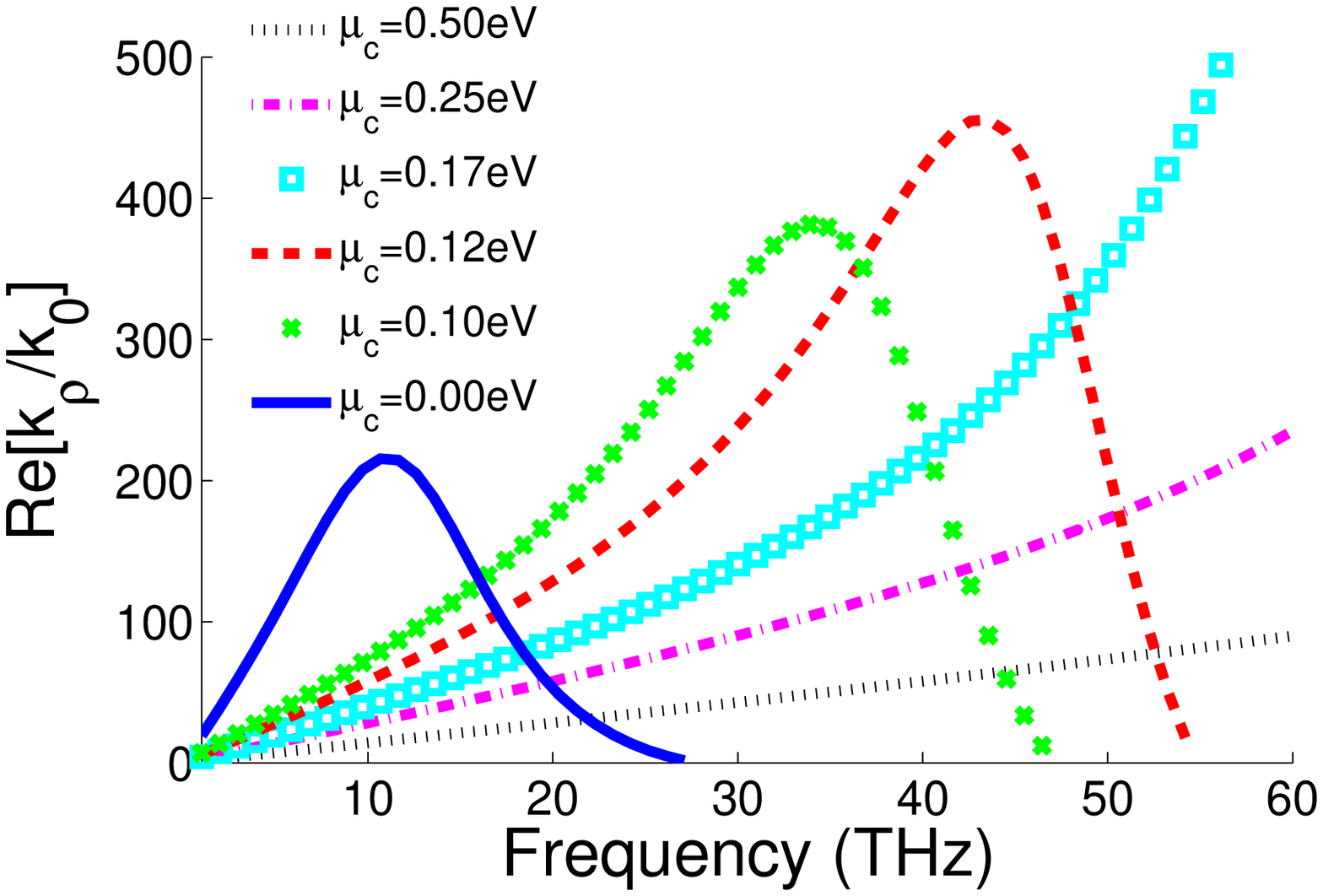}}
\subfloat[]{\label{fig:_Im_kp}
\includegraphics[width=0.45\columnwidth]{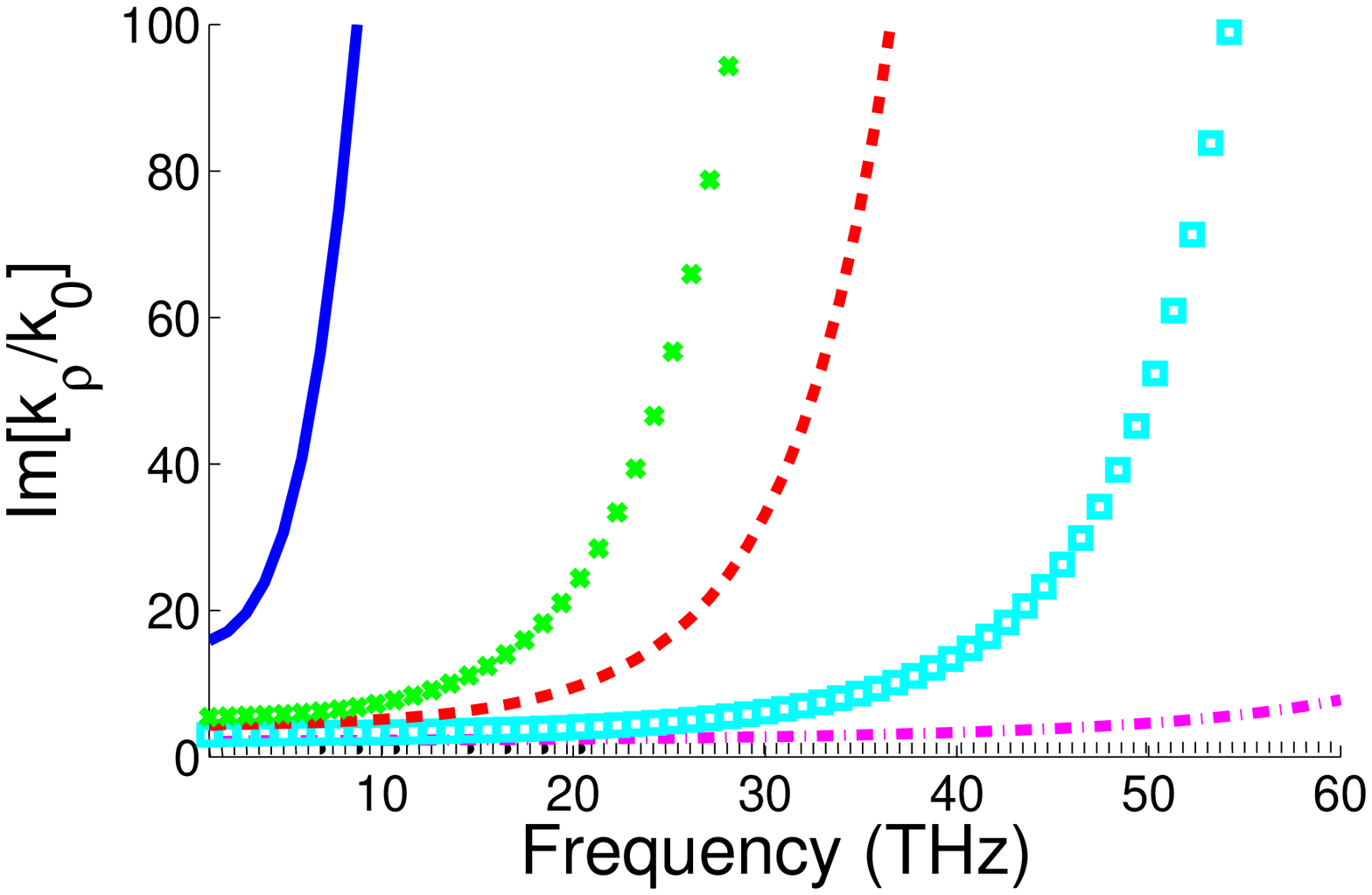}} \\
\subfloat[]{\label{fig:_Z_C}
\includegraphics[width=0.45\columnwidth]{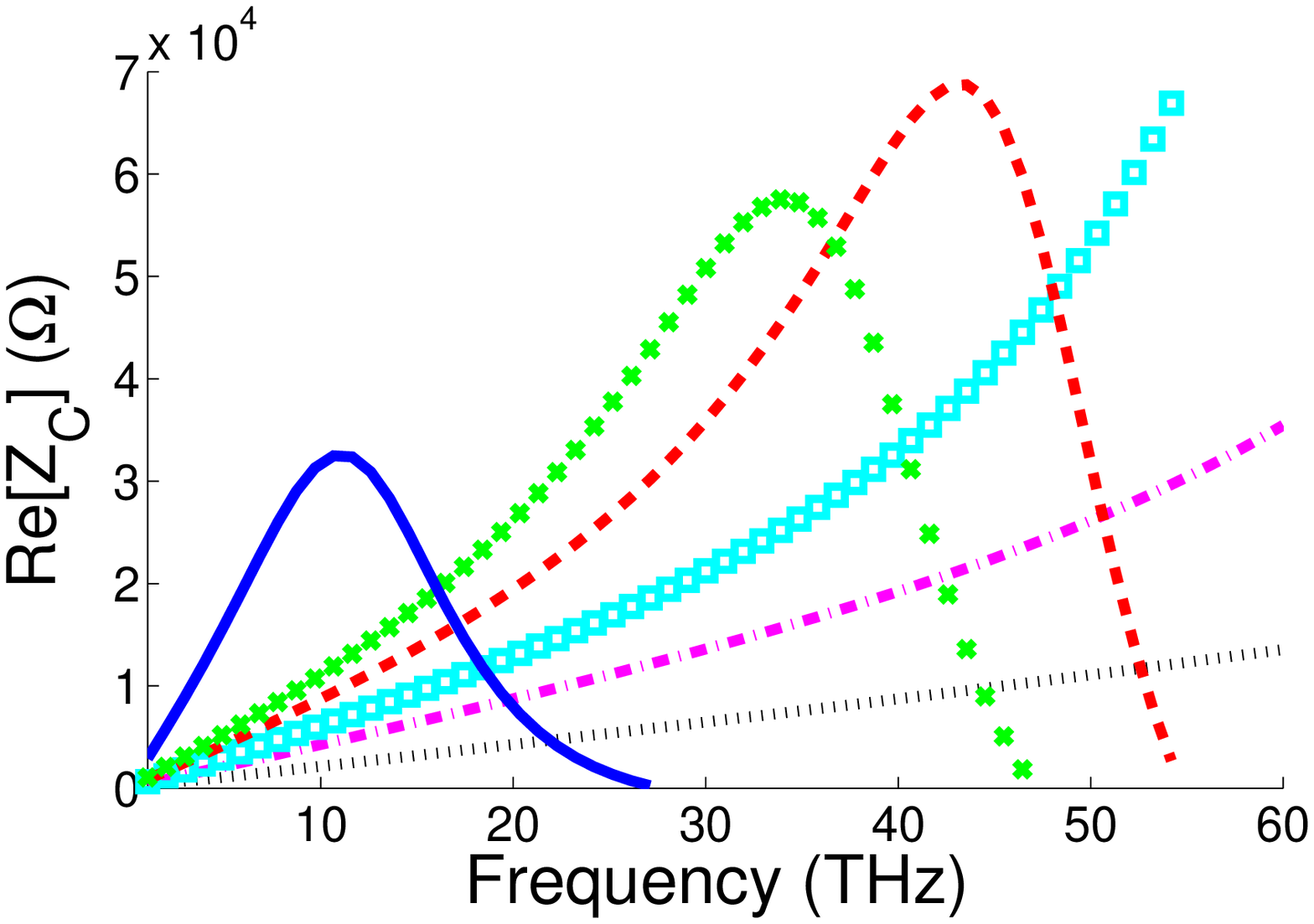}}
\subfloat[]{\label{fig:_Z_C_imag}
\includegraphics[width=0.45\columnwidth]{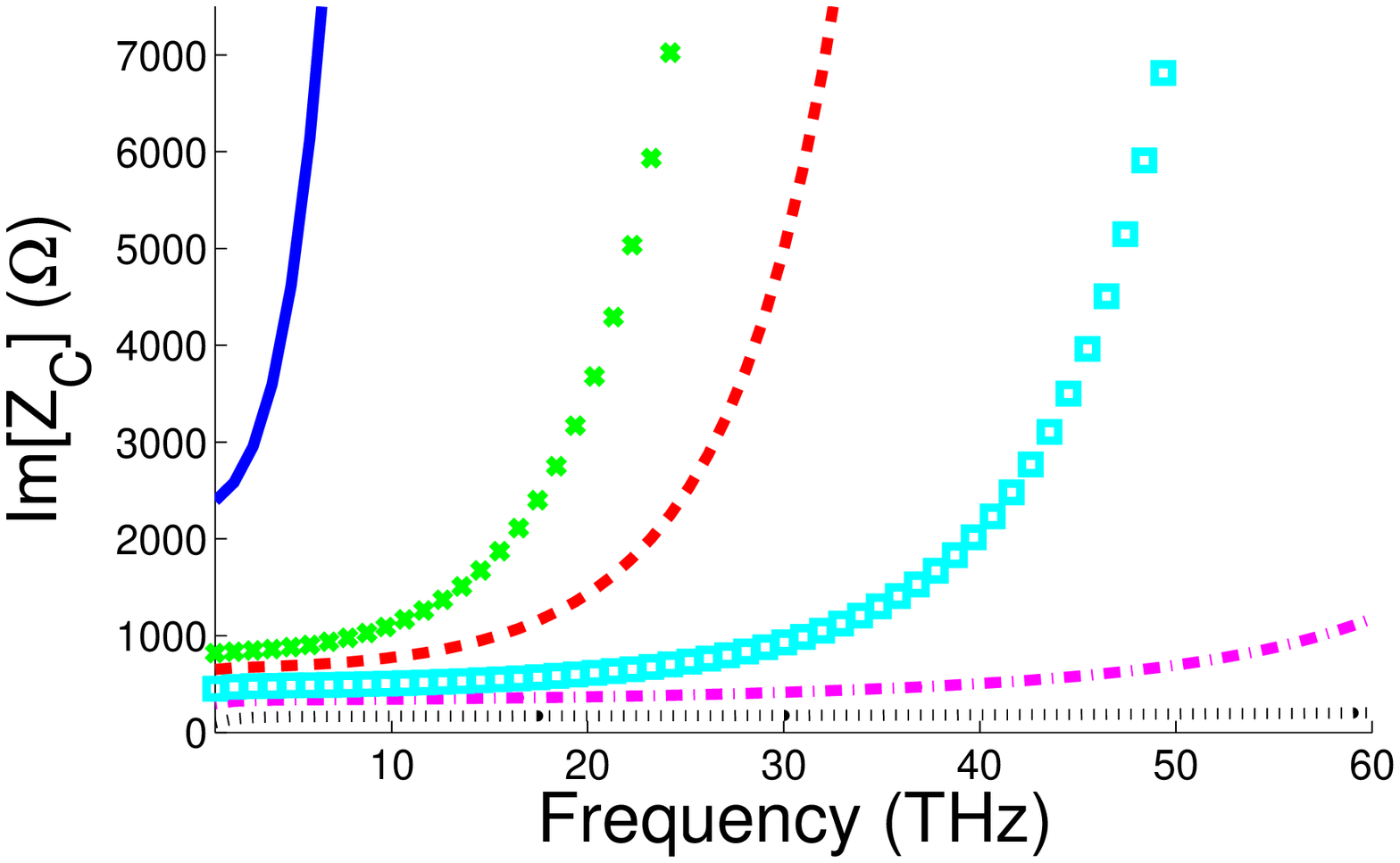}}
\caption{Normalized dispersion relation (a), attenuation constant (b), and real and imaginary components (c-d) of the characteristic impedance of a SPP wave propagating on an air-graphene-dielectric interface versus graphene chemical potential $\mu_c$ computed using Eqs.~(\ref{eq:Dispersion_TM_general}) and (\ref{eq:Characteristic_impedance}). The dielectric permittivity is $\varepsilon_r=4.0$ and graphene parameters are $T=300$~K and $\tau=0.2$~ps.} \label{fig:_Propagation_constant}
\end{figure}

Let us consider, for the sake of illustration, a graphene sheet transferred on a dielectric with permittivity $\varepsilon_r=4$. The parameters of graphene are $\tau=1/(2\Gamma)=0.2$~ps and $T=300$~K, in agreement with measured data \cite{Kim11}. The characteristics of a SPP propagating on the sheet are shown in Fig.~\ref{fig:_Propagation_constant} for different values of graphene chemical potential. We find that the propagation constant and characteristic impedance of the SPP mode can be tuned over a large range by varying $\mu_c$. Focusing for instance in the range between $25$ and $30$ THz, the structure does not support the propagation of SPP when $\mu_c=0.0$~eV (because Im$(\sigma)>0$, as demonstrated in \cite{Hanson08}), it can support SPP propagating with large amount of losses (for instance with $\mu_c=0.1$~eV) or it can even support confined and low-loss propagating modes ($\mu_c=0.5$~eV). This rich variety of propagation characteristics provides unprecedent guiding opportunities in the field of plasmonics at near infrared frequencies, which are exploited below to proposed graphene-strip plasmonic waveguides with switching capabilities.

\subsection{Operation principle of graphene-based switches}
\label{Sec:Operation_principle}

\begin{figure} \centering
\subfloat[]{\label{fig:_Sheet_switch_on}
\includegraphics[width=0.5\columnwidth]{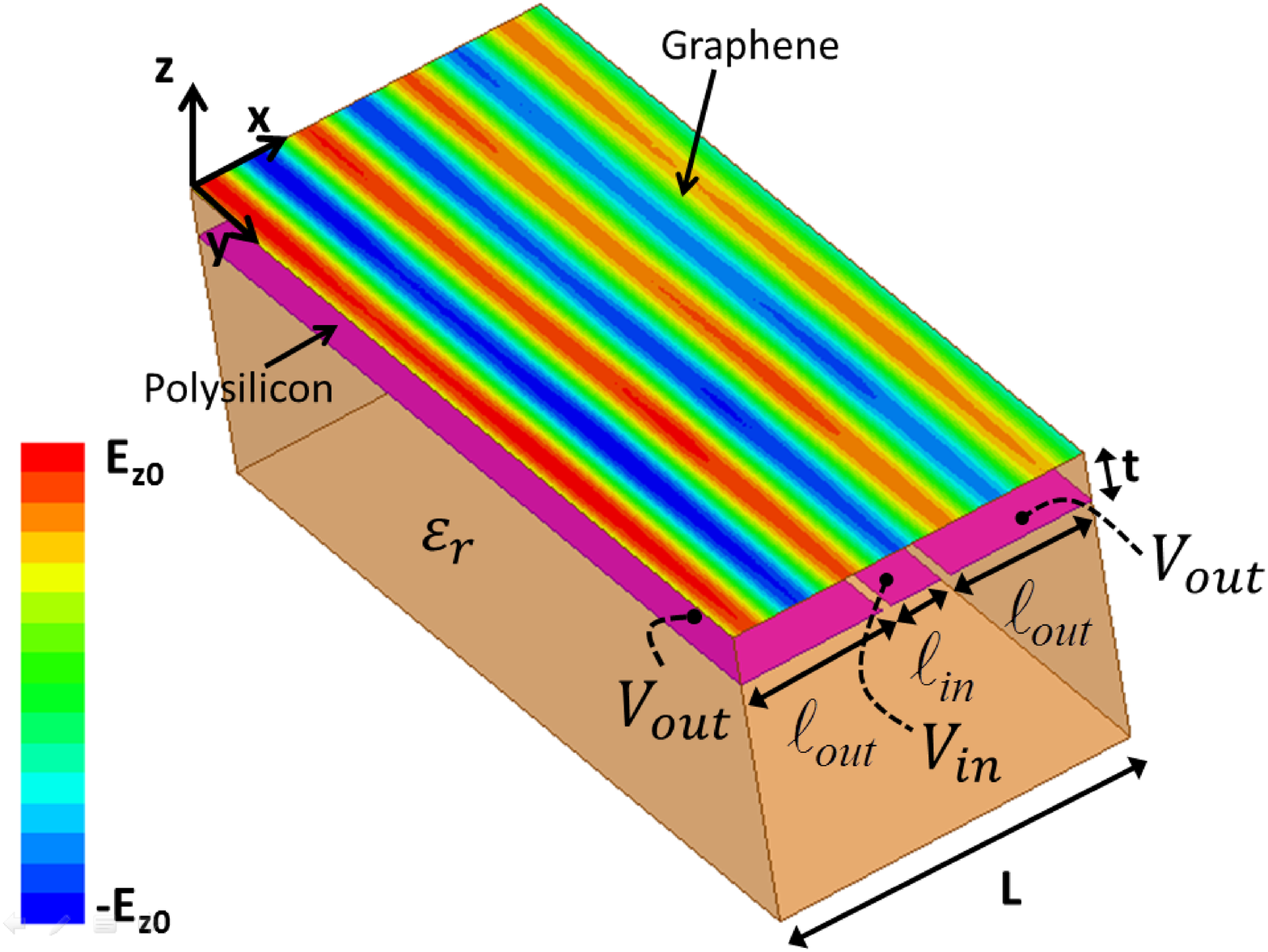}}
\subfloat[]{\label{fig:_Sheet_switch_off}
\includegraphics[width=0.5\columnwidth]{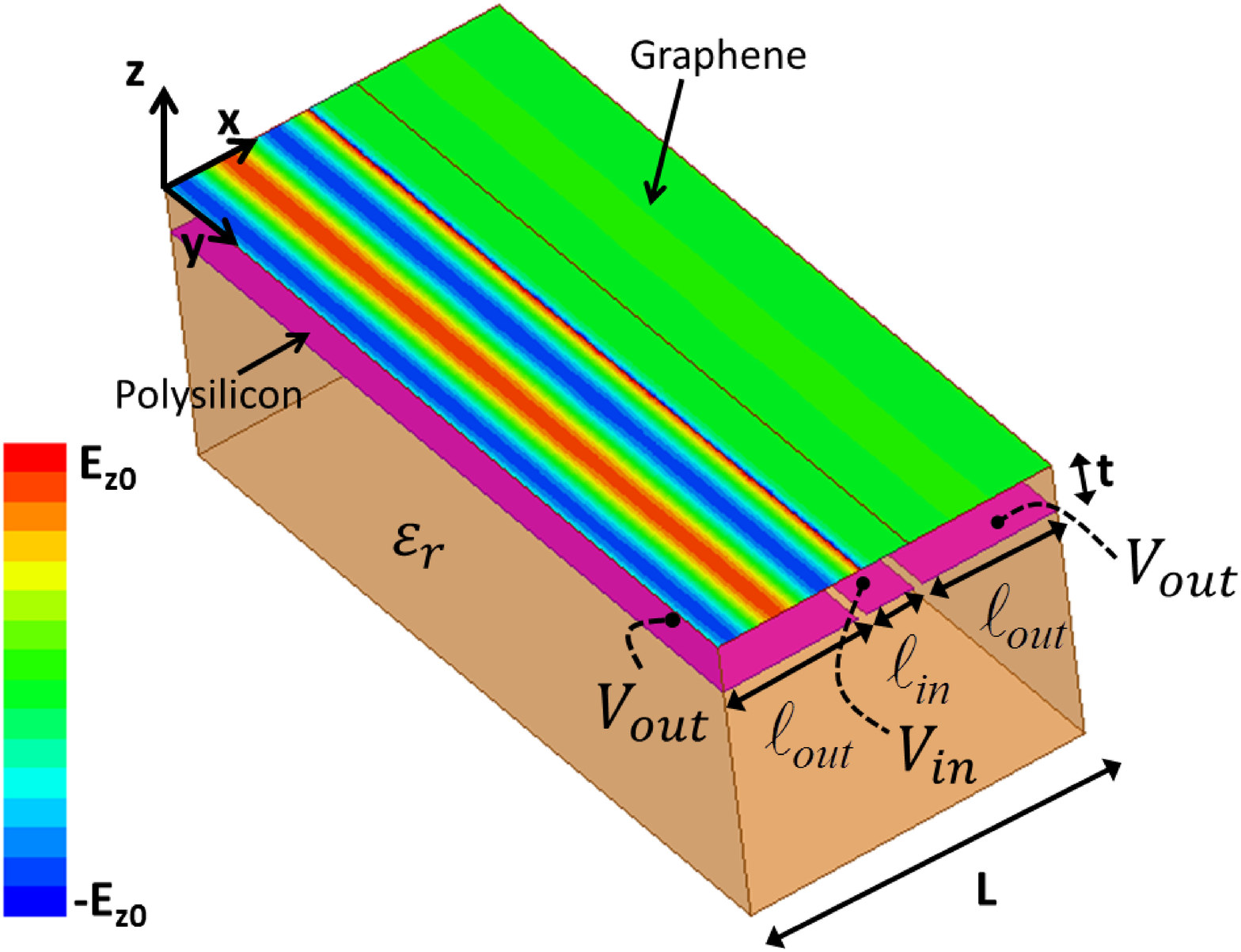}}
\caption{Proposed graphene-based $2$D sheet plasmonic switch. The device comprises a monolayer graphene sheet transferred onto a dielectric ($\varepsilon_r$) and three polysilicon gating pads placed at a distance $t$ below the sheet. The permittivity of the supporting substrate is also set to $\varepsilon_r$. The guiding properties of the SPP propagating along the sheet are controlled via the electric field effect by the DC bias applied to the gating pads. (a) Switch ON. Simulated results showing the $z$ component of the electric field, $E_z$, of a SPP wave propagating along the sheet. The central and outer pads are biased with voltages $V_{out}$ and $V_{in}$, chosen to provide the same chemical potential ($\mu_c=0.5$~eV) to the whole graphene sheet. (b) Switch OFF. Similar to (a) but here $V_{in}$ is chosen to provide a chemical potential of $\mu_c=0.1$~eV to the inner surface of the graphene sheet. The parameters of the structure are $\varepsilon_r=4.0$, $L=350$~nm, $\ell_{in}=50$~nm, $t=20$~nm, $T=300$~K, $\tau=0.2$~ps, and the operation frequency is set to $28$~THz.} \label{fig:_Sheet_switch}
\end{figure}
\begin{figure} \centering
\subfloat[]{\label{fig:_TL_switch_on}
\includegraphics[width=0.5\columnwidth]{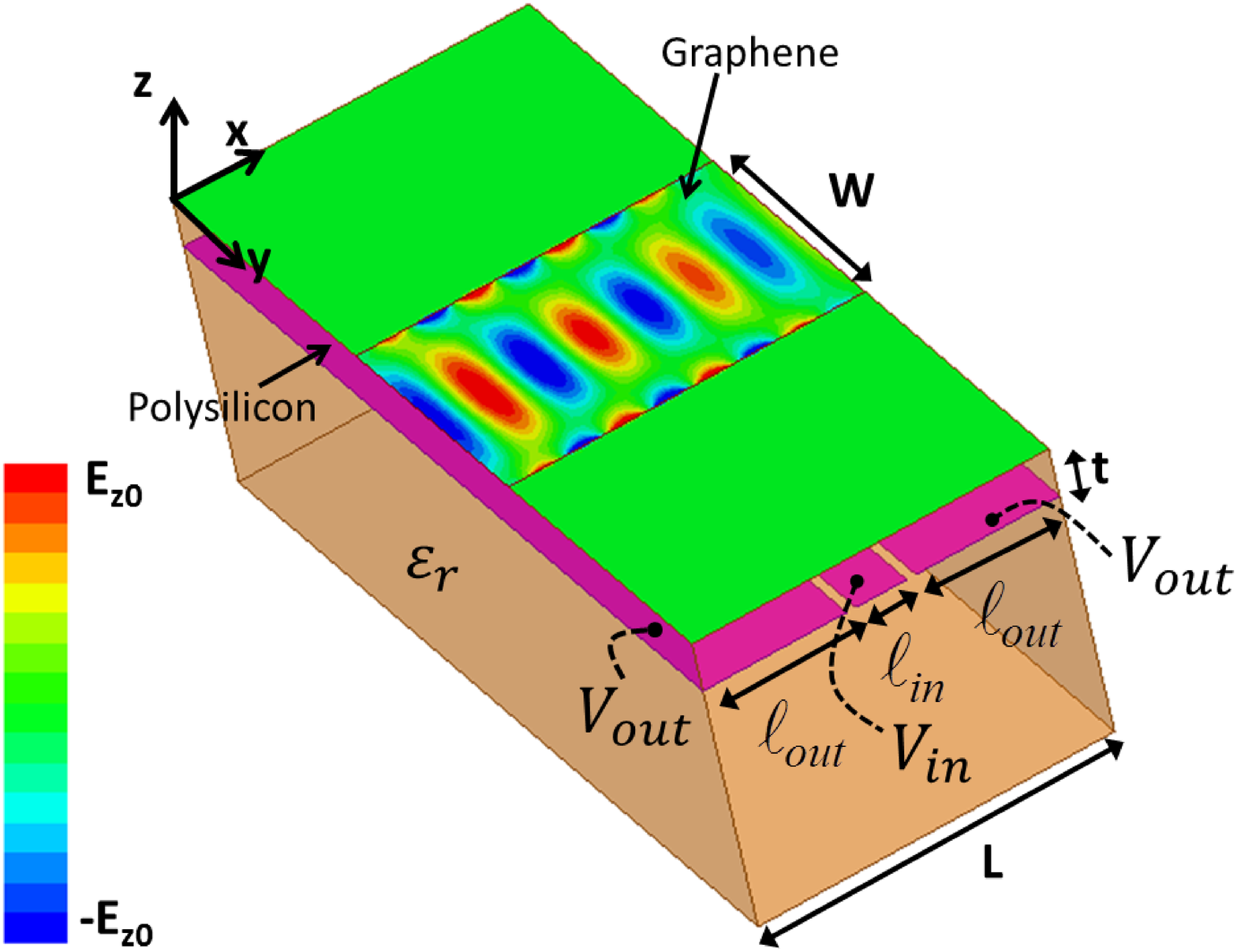}}
\subfloat[]{\label{fig:_TL_switch_off}
\includegraphics[width=0.5\columnwidth]{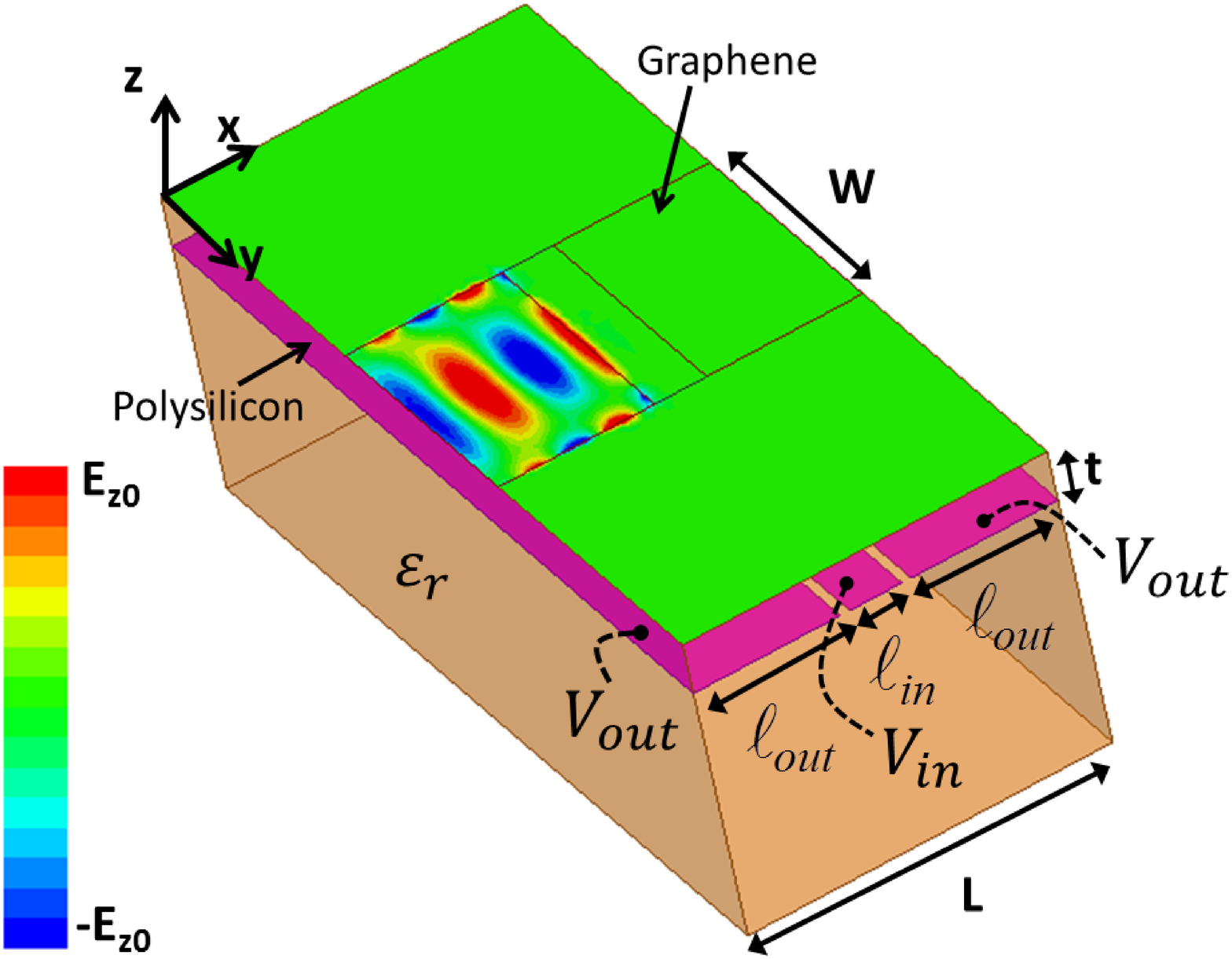}}
\caption{Proposed graphene-based strip plasmonic switch. The device is similar to the switch shown in Fig.~\ref{fig:_Sheet_switch}, but here the graphene sheet is replaced by a strip of width $W$. (a) Switch ON. Simulated results showing the $z$ component of the electric field, $E_z$, of a SPP wave propagating along the strip. The voltages $V_{out}$ and $V_{in}$ are chosen to provide the same chemical potential ($\mu_c=0.5$~eV) to the whole graphene strip. (b) Switch OFF. Similar to (a) but here $V_{in}$ is chosen to provide a chemical potential of $\mu_c=0.1$~eV to the inner section of the strip. The parameters of the structure are $\varepsilon_r=4.0$, $L=350$~nm, $W=150$~nm, $\ell_{in}=50$~nm, $t=20$~nm, $T=300$~K, $\tau=0.2$~ps, and the operation frequency is set to $28$~THz.} \label{fig:_Strip_switch}
\end{figure}

The proposed graphene-based switches are shown in Fig.~\ref{fig:_Sheet_switch} and in Fig.~\ref{fig:_Strip_switch}. The electric fields of the structures are computed using the commercial software HFSS, as detailed in the next section. The devices are composed of a host graphene waveguide, namely a $2$D sheet in Fig.~\ref{fig:_Sheet_switch} and a finite strip in Fig.~\ref{fig:_Strip_switch}, transferred on a dielectric (with $\varepsilon_r$) and of three polysilicon gating pads beneath the waveguide. The permittivity of the supporting substrate is also set to $\varepsilon_r$. The switch is located in the inner section of the waveguide, above the central pad, and the outer sections of the waveguide connect the switch to the input and output ports of the device. The characteristics of the SPP propagating on the switch and on the other sections of the waveguides are controlled by the DC voltage applied to the gating pads. Specifically, the outer pads are biased with a voltage $V_{out}$, which provides to the graphene area located above a high chemical potential, whereas the central pad is biased with a voltage $V_{in}$. The operation principle of the switch is as follows. In the ON state [see Fig.~\ref{fig:_Sheet_switch}(a) and Fig.~\ref{fig:_Strip_switch}(a)], the DC voltages $V_{out}$ and $V_{in}$ are chosen to provide the same chemical potential to the whole graphene waveguide. In this state, the device behaves as a simple transmission line able to propagate an incoming wave from the input to the output port. In the OFF state [see Fig.~\ref{fig:_Sheet_switch}(b) and Fig.~\ref{fig:_Strip_switch}(b)], the voltage $V_{in}$ is modified to provide a much lower value to the chemical potential of the graphene located in the switch area. In this state, an incoming wave propagating on the waveguide finds a strong discontinuity due to the different characteristic impedance and propagation constant of the central section of the line. Specifically, there is a wave reflected back to the input port due to the large mismatch between the different regions of the waveguide, some energy is dissipated in the switch due to the large losses that now arise, and a highly attenuated wave is transmitted towards the output port. The performance of the switch is determined by the isolation between the input and output ports at the OFF state and by the insertion loss in the ON state. It depends on the length of the switch ($\ell_{in}$) and on the range of chemical potential values achievable by graphene, as will be discussed in detail in \secref{Sec:Numerical_results}.

\begin{figure} \centering
\subfloat[]{\label{fig:_Fabrication_case_1}
\includegraphics[width=0.45\columnwidth]{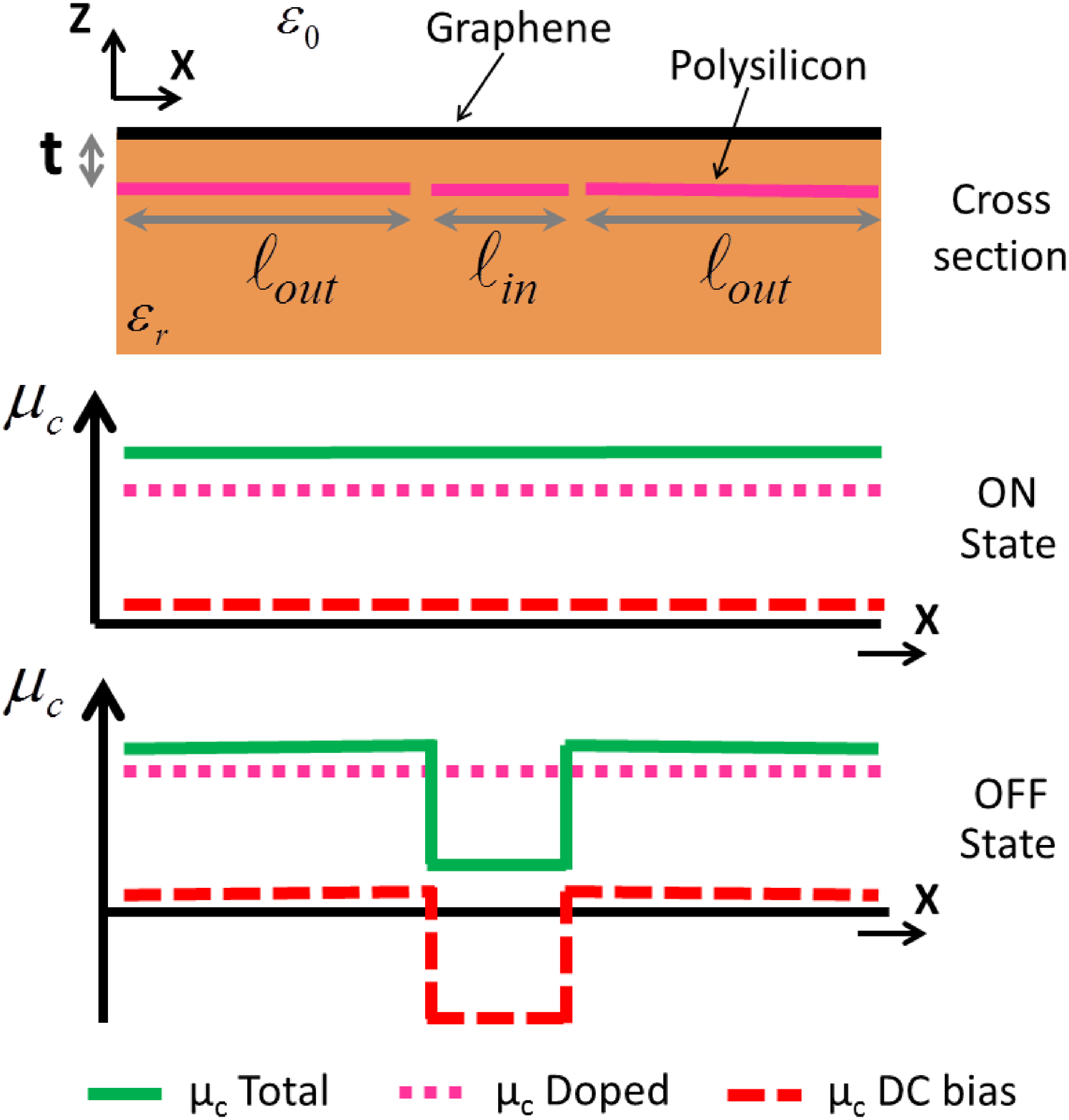}}
\subfloat[]{\label{fig:_Fabrication_case_2}
\includegraphics[width=0.45\columnwidth]{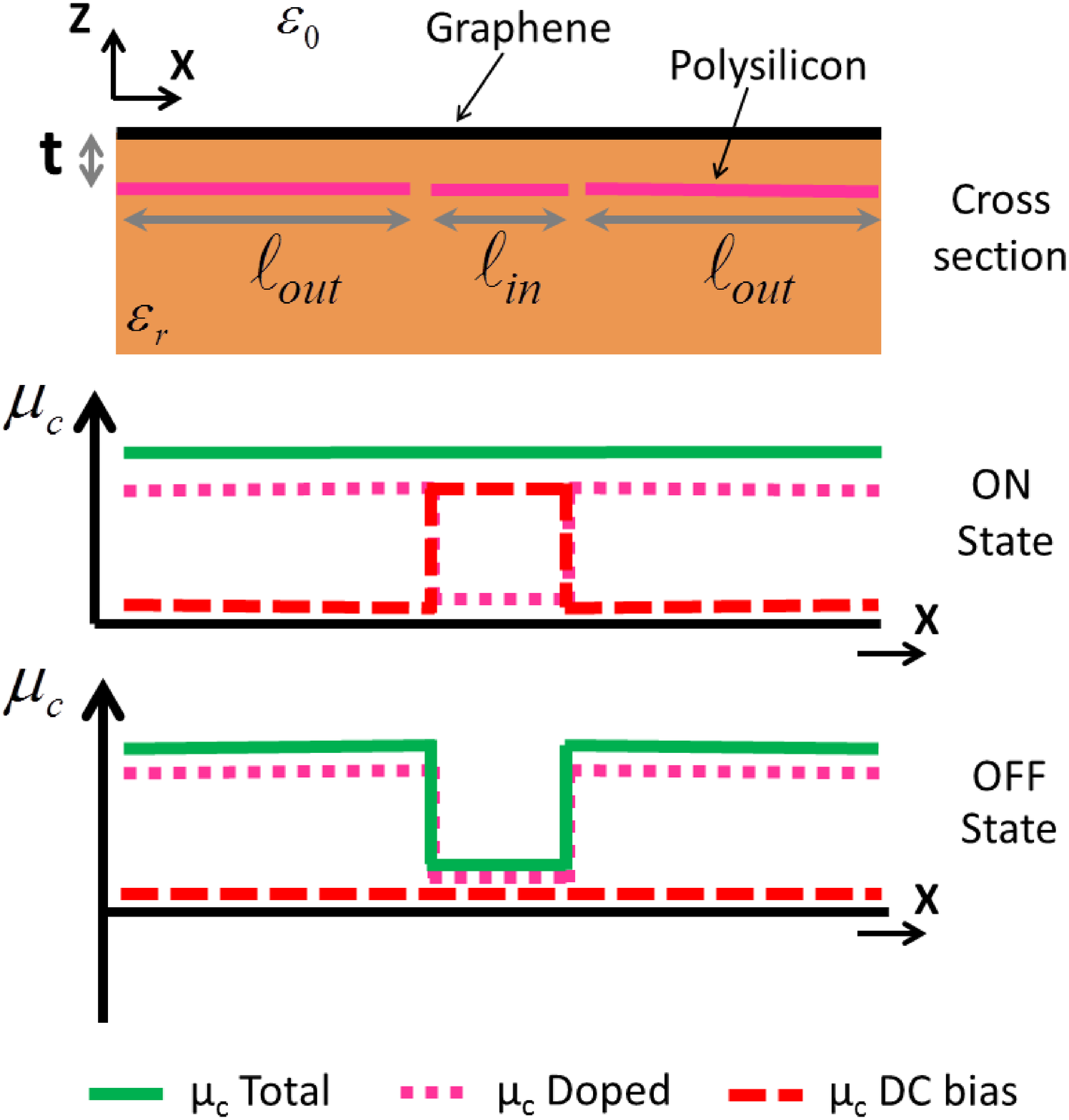}}
\caption{Cross section of the proposed switch and chemical potential profile along the `x' axis of the graphene area for the ON and OFF states of the device. The different contributions to the chemical potential of graphene (solid line), namely chemical doping (dotted line) and elecrostatic DC bias (dashed line), are also shown. (a) Uniformly highly chemically doped graphene. The OFF state is obtained by applying a negative DC bias to the central gating pad. (b) Non-uniformly chemically doped graphene. Outer and inner surfaces of graphene are highly and slightly chemically doped, respectively. The ON state is obtained by applying a positive DC bias to the central gating pad.}\label{fig:_Fabrication}
\end{figure}

We propose two different alternatives, illustrated in \figref{fig:_Fabrication}, for the technological implementation of the switches. In both cases, the ON state is obtained by providing high chemical potential to the whole graphene surface while in the OFF state the chemical potential of the central section is highly reduced. The first approach, shown in \figref{fig:_Fabrication}(a), uses uniformly highly doped graphene sheets/strips. Note that recent fabrication techniques have demonstrated large control of graphene chemical doping \cite{Wang09,Kim09_large_scale_graphene,Reina09}, and measured values around $0.4$~eV have already been reported \cite{Bae10}. In this case, the ON state is obtained by applying a low DC voltage ($V_{out}=V_{in}=V_{L}$) to the gating pads, which provide to the whole graphene area the required chemical potential. On the other hand, the OFF state is obtained by applying a negative voltage ($-V_H$) to the central gating pad. Indeed, due to the ambipolarity property of graphene \cite{Geim2007, Geim2009}, an applied negative DC voltage decreases the chemical potential of graphene. Therefore, the central and outer gating pads are biased with voltages $V_{L}$ and $-V_{H}$, respectively. The second approach [see \figref{fig:_Fabrication}(b)] relays on tailoring the chemical doping of the different graphene regions. In this way, the outer graphene surfaces are highly chemically doped, whereas the inner surface is slightly doped. The ON state is obtained providing a low DC bias voltage to the outer gating pads ($V_{out}=V_L$) and a larger bias to the central one ($V_{in}=V_H$), while in the OFF state the voltage applied to the central pad is reduced ($V_{in}=V_{L}$). Note that this second approach requires a more complicated fabrication process due to the tailoring of the chemical doping applied to the graphene area.
\subsection{Electromagnetic modeling}
\label{Sec:Graphene_electromagnetic_modeling}
The electromagnetic modeling of the proposed graphene-based switches is performed using two different techniques, namely a TL formalism combined with an ABCD transfer-matrix approach \cite{Pozar05} and a commercial full-wave software (HFSS) \cite{HFSS12} based on the finite element method.

\begin{figure} \centering
\includegraphics[width=0.6\columnwidth]{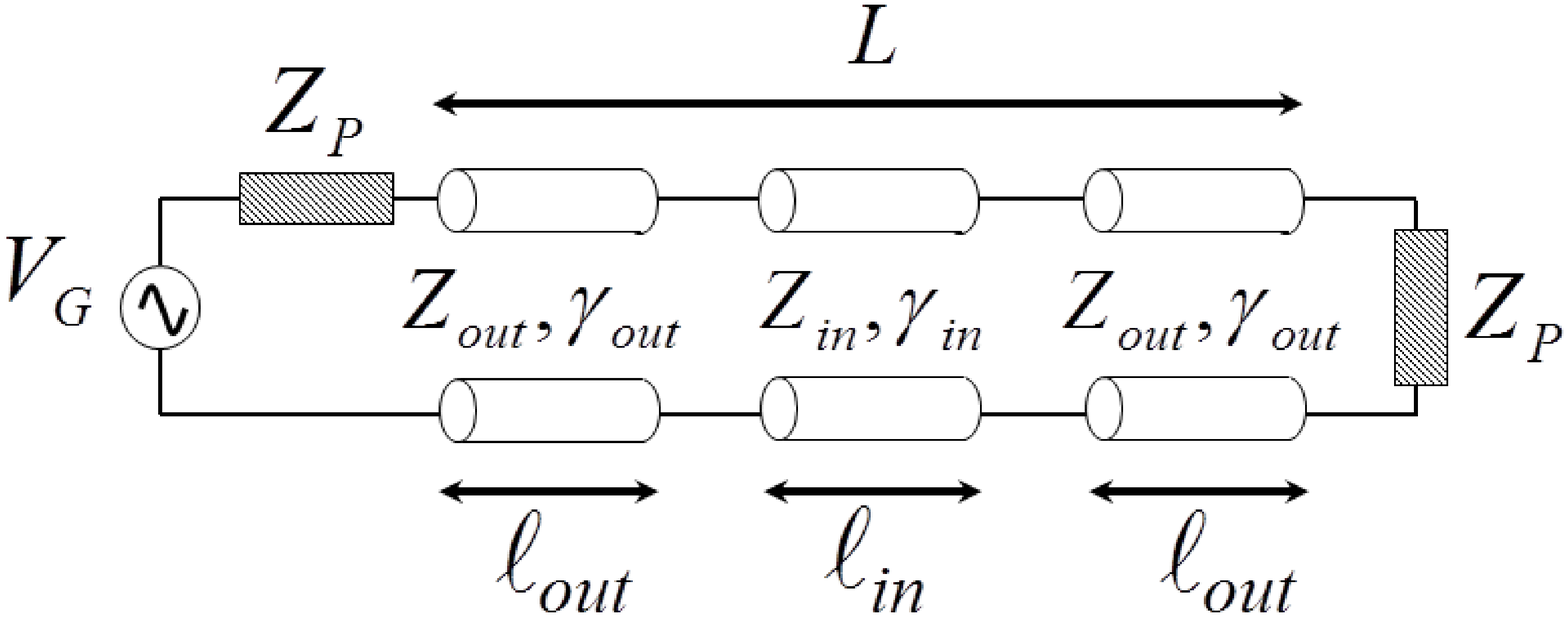}
\caption{Equivalent transmission line model of the proposed graphene-based switches shown in Fig.~\ref{fig:_Sheet_switch} and in Fig.~\ref{fig:_Strip_switch}.} \label{fig:_TL_model_switch}
\end{figure}

The proposed graphene-based waveguide switch can be modeled by cascading three transmission lines, as shown in Fig.~\ref{fig:_TL_model_switch}. Each TL corresponds to a section of the host waveguide. The outer lines, characterized by a propagation constant $\gamma_{out}$ and characteristic impedance $Z_{out}$, are related to the outer sections of the device. In addition, the switch (inner region of the waveguide) is modeled by the central TL, which present a propagation constant and characteristic impedance of $\gamma_{in}$ and $Z_{in}$, respectively. The different lines are combined using an ABCD matrix-transfer approach, and the corresponding scattering parameters are then recovered using standard techniques \cite{Pozar05}. For simplicity, scattering parameters are refereed to the port impedance $Z_P$ which is set equal to the characteristic impedance of the outer waveguide sections, $Z_{out}$. However, it is important to point out that this method is approximate. It is well-known from transmission line theory \cite{Pozar05, collin91} that this approach only considers the fundamental mode that propagates along the waveguide, and neglects the presence of higher order modes that are excited at the discontinuity between different sections. Moreover, note that the excitation and influence of these higher order modes increase when i) the guiding characteristics of the different waveguide sections are very different, and ii) the length of one of these sections is very small \cite{collin91}. Nevertheless, this approximate approach is indeed useful: it provides extremely fast preliminary results and physical insight into the operation principle of the proposed switches.

In the case of the graphene-based $2$D sheet switch shown in \figref{fig:_Sheet_switch}, the propagating constant and characteristic impedance of the equivalent transmission lines are directly obtained using Eqs.~(\ref{eq:Dispersion_TM_general}) and (\ref{eq:Characteristic_impedance}), respectively. In the case of graphene-based strip switches, depicted in \figref{fig:_Strip_switch}, no analytical formulas are available. Here, we employ full-wave simulations to extract the propagation constant of SPP propagating on various infinitely-long graphene strips. Then, the extracted wavenumbers are included in the TL formalism to characterize the complete switch.

The second approach is based on simulating the complete switch in a full-wave simulator. To this purpose, we have employed the commercial package HFSS \cite{HFSS12}, which implements the finite element method (FEM) \cite{jin93, volakis98}. There, graphene surfaces are modeled as infinitesimally thin sheets/strips, where surface impedance boundary conditions ($Z_{Suf}=1/\sigma$) are imposed. We have verified that $3$D numerical FEM simulations are able to accurately model plasmon propagation on graphene waveguides. However, we have observed some discrepancies between the plasmons characteristics -namely propagation constant and characteristic impedance- obtained by the $2$D numerical waveport solutions and the expected ones. These discrepancies may arise due to numerical instabilities related to the large surface reactance of the infinitesimally thin graphene. Consequently, though we are able to accurately compute the scattering parameters of the structure under analysis, they are referred to an unknown port impedance ($Z_P$). In order to solve this issue, we employ a simpler structure with exactly the same waveport configuration to rigorously extract the port impedance. This is done by simulating a uniform graphene-based sheet/strip waveguide with known propagation constant and characteristic impedance (see Eq.~(\ref{eq:Dispersion_TM_general}) and \cite{Nikitin11}). Then, $Z_P$ is analytically retrieved from the resulting scattering parameters applying standard techniques \cite{Pozar05, collin91}. Finally, this impedance is employed to renormalize the scattering parameters of the initial complex structure to any desired impedance, $Z_P=Z_{out}$ in our case.
\section{Numerical results}
\label{Sec:Numerical_results}
In this section, we investigate the characteristics of the proposed switches in terms of their scattering parameters. These parameters are commonly employed in the microwaves and terahertz frequency ranges \cite{Pozar05} and are perfectly suited to evaluate the behavior of a switch. Note that an ideal switch in its ON state propagates all input energy towards the output port (i.e. $S_{21}\approx1$), while no energy is transmitted in its OFF state (i.e. $S_{21}\approx0$). First, we verify that the two numerical techniques employed to characterize plasmon propagation on graphene waveguides, namely the TL approach and the commercial software HFSS, leads to similar results. Then, we focus for simplicity on switches suspended on free space. There, we study the characteristics of the switches in their ON and OFF states, and we present a parametric study of the switches performance as a function of their features. Finally, we extend this analysis to consider realistic graphene-based switches, taking into account the presence of a substrate. In our study, we consider a relaxation time of $0.2$~ps, in agreement with measured values of graphene carrier mobility \cite{Kim11}, and a temperature of $T=300^{\circ}$~K (room temperature). For simplicity, we neglect the possible fluctuations of graphene relaxation time due to optical phonons \cite{Jablan09}.

\begin{figure} \centering
\includegraphics[width=0.6\columnwidth]{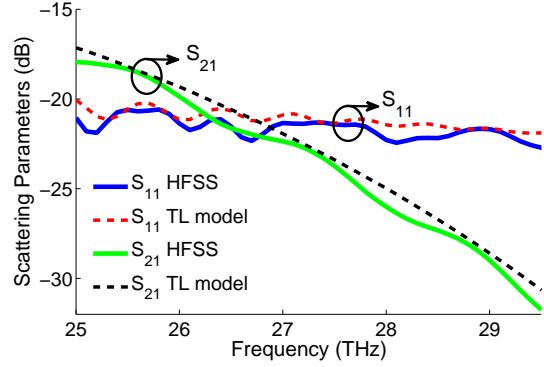}
\caption{Scattering parameters of the structure shown in Fig.~\ref{fig:_Sheet_switch}, with $\varepsilon_r=1$, $L=3~\mu$m and $\ell_{in}=1~\mu$m, computed using the transmission line approach and the commercial software HFSS. The chemical potential of the outer and central graphene waveguide sections are set to  $\mu_{c_{out}}=0.2$~eV and $\mu_{c_{in}}=0.15$~eV.
} \label{fig:_Intermediate_case}
\end{figure}

In order to assess the accuracy of the numerical methods employed to study graphene-based switches, we first consider the propagation of surface plasmons on the structure shown in \figref{fig:_Sheet_switch}. The parameters of the structure are $\varepsilon_r=1$, $L=3~\mu$m and $\ell_{in}=1~\mu$m, and the chemical potential of the central and outer graphene waveguide sections is set to $0.2$~eV and $0.15$~eV, respectively. The scattering parameters of the structure, computed using the TL approach and HFSS, are shown in \figref{fig:_Intermediate_case}. Very good agreement is found between the methods, verifying that the propagation of surface plasmons on the graphene waveguide is correctly modeled. Note that the accurate results provided by the TL approach are due to the negligible influence of higher order modes in this structure, which are barely excited in the weak discontinuity between the different waveguide sections.

\begin{figure} \centering
\subfloat[]{\label{fig:_S_param_sheet_fs}
\includegraphics[width=0.5\columnwidth]{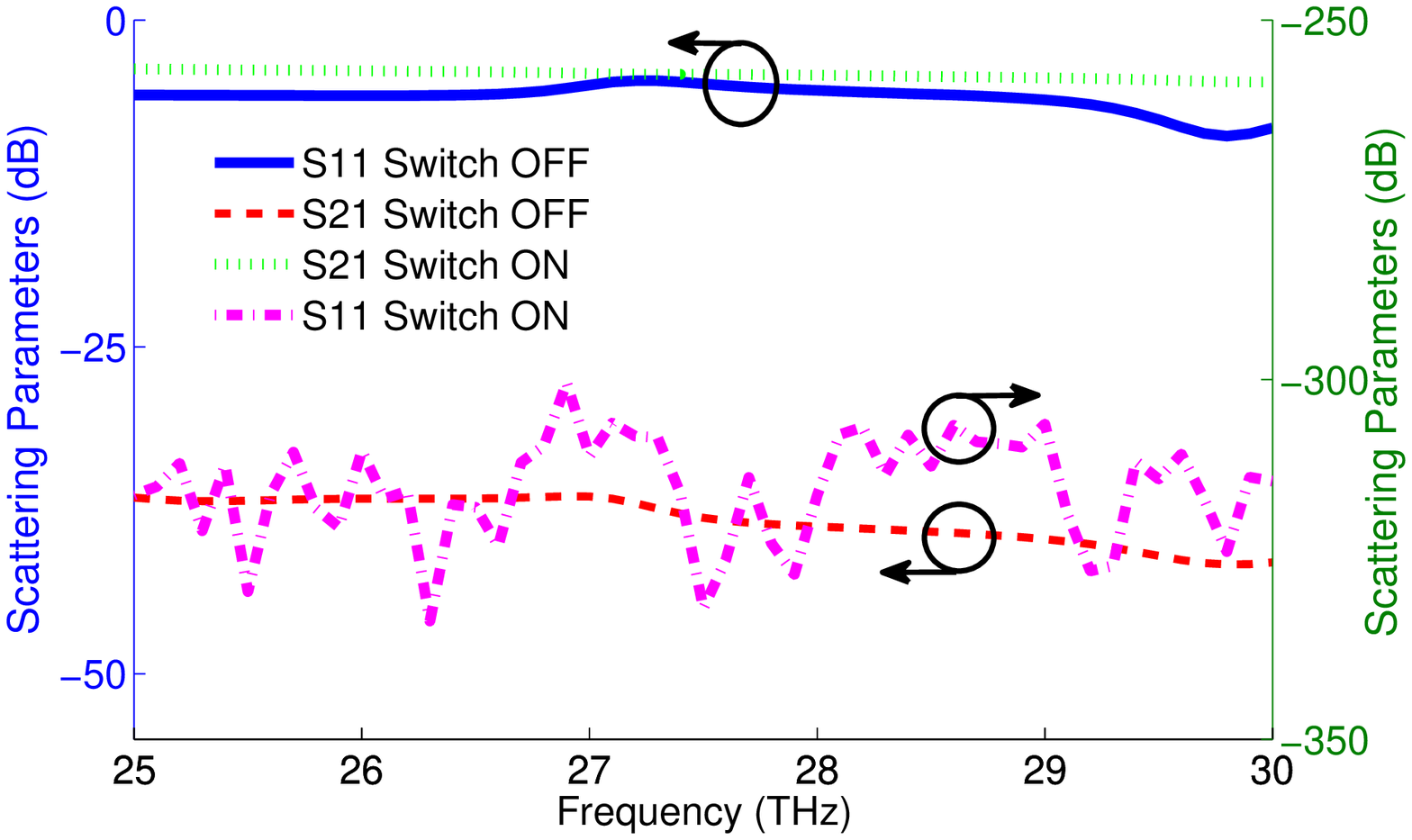}}
\subfloat[]{\label{fig:_S_param_strip_fs}
\includegraphics[width=0.5\columnwidth]{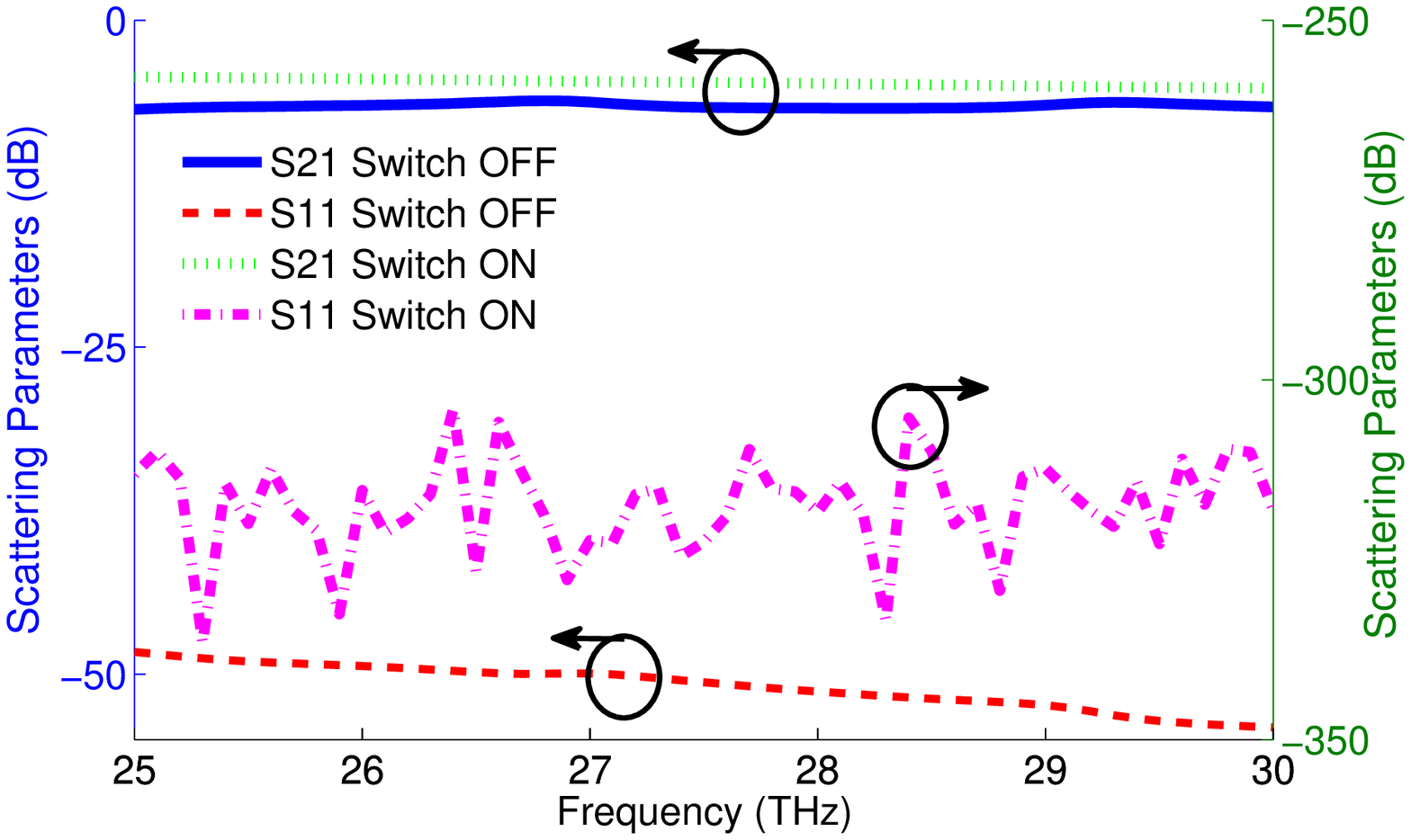}}
\caption{Simulated scattering parameters of the proposed graphene-based switches, suspended in free-space, at their ON and OFF states. The parameters of the device are $L=1.75~\mu$m and $\ell_{in}=0.5~\mu$m. (a) Graphene-based $2$D sheet switch, see Fig.~\ref{fig:_Sheet_switch}. (b) Graphene-based strip switch with $W=0.2~\mu$m, see Fig.~\ref{fig:_Strip_switch}.}\label{fig:_S_param_sheet}
\end{figure}

\figref{fig:_S_param_sheet} presents the scattering parameters, obtained with the commercial software HFSS, of the proposed graphene-based switches suspended in free-space (see Fig.~\ref{fig:_Sheet_switch} and  Fig.~\ref{fig:_Strip_switch}, with $\varepsilon_r=1$). The length of the total device and the switch section are set to $L=1.75~\mu$m and $\ell_{in}=0.5~\mu$m. In the ON state, a DC bias voltage is applied to all gating pads to provide a chemical potential of $\mu_c=0.5$~eV to the whole graphene area, while in the OFF state the DC bias applied to the central pad is reduced, leading to a chemical potential of $0.1$~eV for the switch area. Figures \ref{fig:_S_param_sheet}(a)-\ref{fig:_S_param_sheet}(b) show the performance of the proposed graphene-based $2$D sheet and strip switches. In the ON state, the switches behave as a transmission line and the input energy propagates towards the output port (i.e. $S_{21}\approx1$). Note that the extremely low value of $S_{11}$, which indicates that the device is very well matched, has been obtained by renormalizing the scattering parameters to the characteristic impedance of the outer waveguide sections ($Z_{out}$, see \secref{Sec:Graphene_electromagnetic_modeling}). This state also provides some dissipation losses, about $4$~dB in both switches, which are due to the intrinsic characteristics of graphene and directly depends on the total length of the device. In the OFF state, the graphene-based $2$D surface and strip switches provide large isolation in the whole frequency band, around $37$ and $50$~dB respectively. Importantly, the use of realistic graphene strips, instead of ideal $2$D sheets, leads to devices with higher isolation levels. This is due to the field confinement of surface plasmons propagating on strips, which is much larger than in case of $2$D sheets \cite{Nikitin11, Christensen12}.

\begin{figure} \centering
\includegraphics[width=0.6\columnwidth]{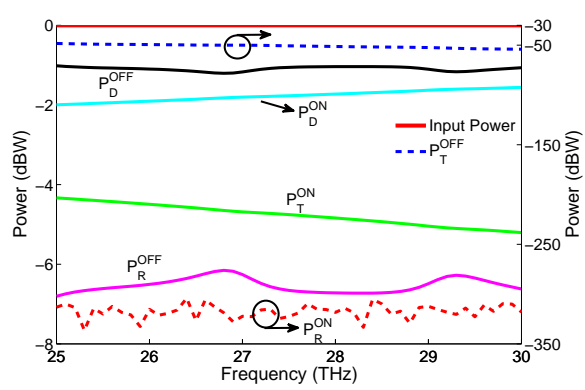}
\caption{Power transmitted, reflected, and dissipated in the graphene-based strip plasmonic switch shown in Fig. 7(b). The superscripts ON and OFF are related to the operation state of the switch, and the subscripts $T$, $R$, and $D$ refer to the power transmitted towards the output port, reflected into the input port, and dissipated in the structure, respectively.} \label{fig:_Power_flow}
\end{figure}

In order to investigate the different power waves flowing along the proposed switches, we present in Fig.~\ref{fig:_Power_flow} the power transmitted, reflected and dissipated in the graphene-based strip plasmonic switch of Fig.~\ref{fig:_S_param_sheet}(b) when the device is fed by a $1$W-power ($0$~dBW) input wave. These quantities can be computed using the the relationship between scattering parameters and power waves \cite{Pozar05}
\begin{align}
P_{T}&=|S_{21}|^2, \label{Power_relationT}\\
P_{R}&=|S_{11}|^2,  \label{Power_relationR}  \\
P_{D}&=1-|S_{11}|^2-|S_{21}|^2, \label{Power_relationD}
\end{align}
where $P_{T}$, $P_{R}$, and $P_{D}$ are related to the total power transmitted towards the output port, reflected into the input port, and dissipated in the structure, respectively. In the ON state all energy propagates into the structure and there are not reflected waves thanks to the good matching of the device. Besides, although there are important dissipation losses, around $-5$~dBW of power is available at the output port. In the OFF state, the propagation is highly attenuated and less than $-50$dBW of power is transmitted towards the output port. As expected, this high attenuation is due to the combination of a reflected wave which is guided back into the input port and to the increase of the dissipation losses at the central waveguide section. In fact, switching from the ON to the OFF state of the switch increases the dissipated power in less than $1$~dBW while the power reflected back increases in more than $300$dBW. This study allows to clearly identify the reflected waves which arise due to the different characteristic impedance of the waveguide sections as the main mechanism which provides the high isolation of the switches.
\begin{figure} \centering
\subfloat[]{\label{fig:_L_S21_Sheet_FS}
\includegraphics[width=0.5\columnwidth]{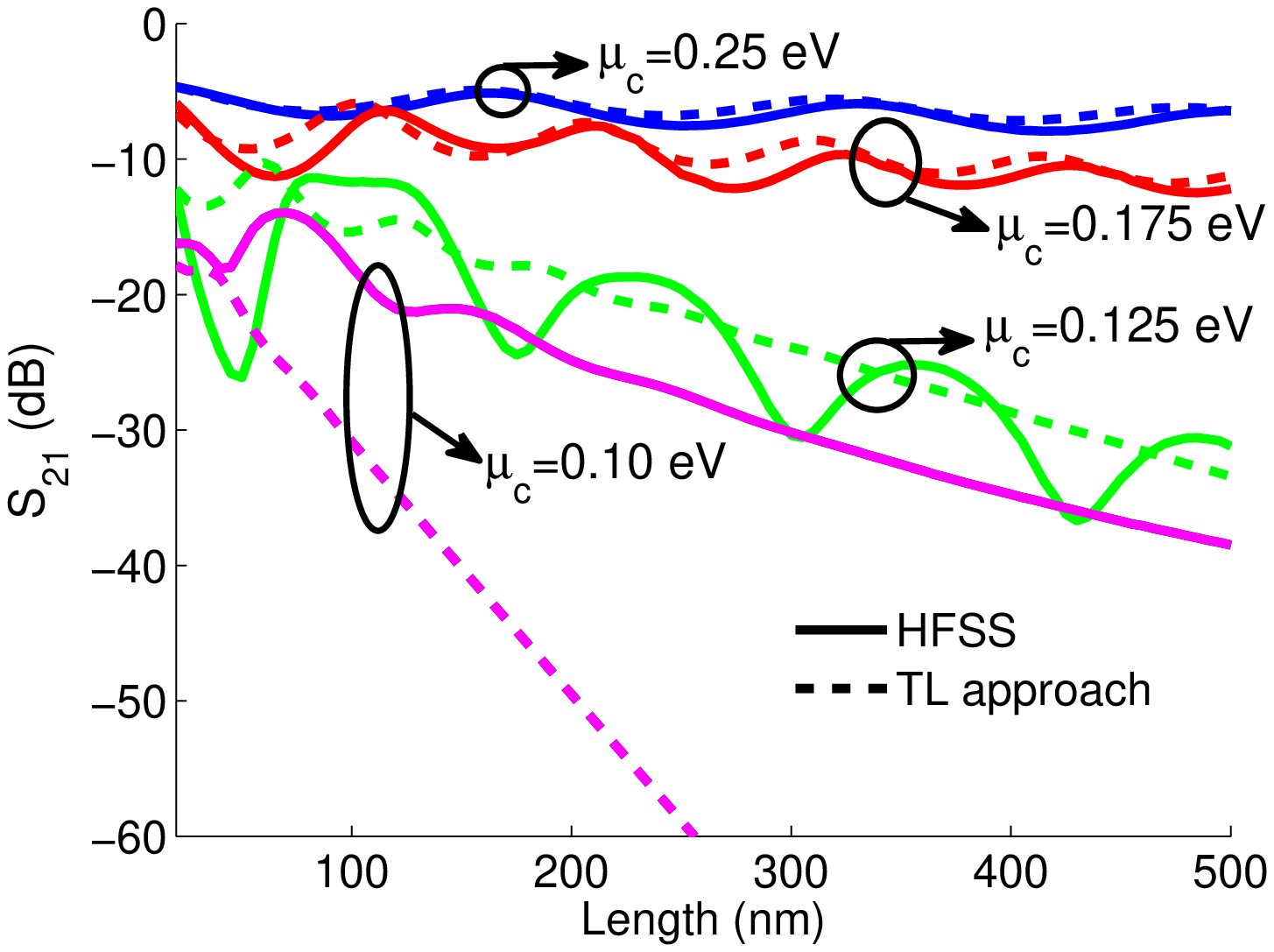}}
\subfloat[]{\label{fig:_L_S21_Strip_FS}
\includegraphics[width=0.5\columnwidth]{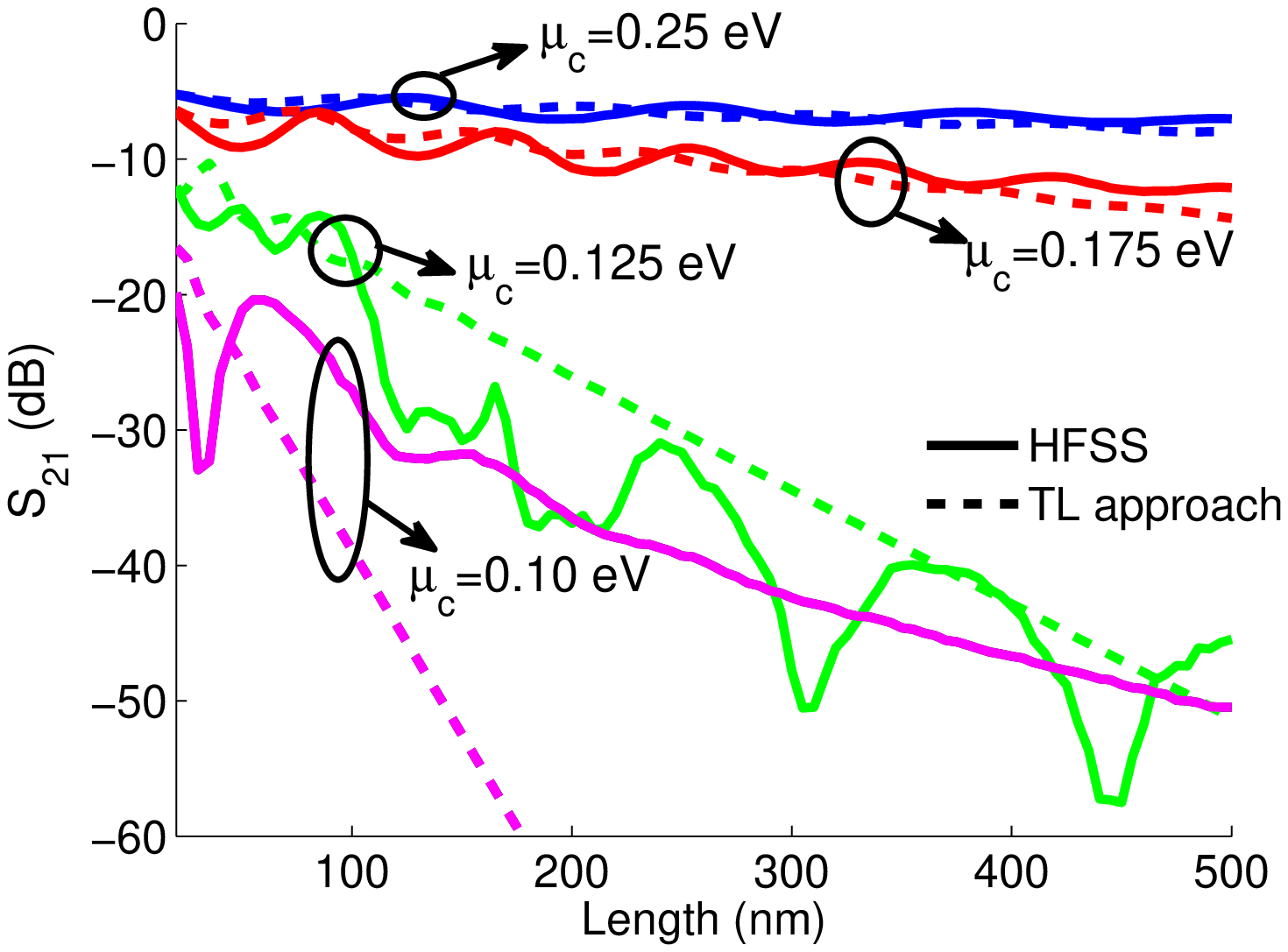}}
\caption{Parametric study of the isolation ($S_{21}$) provided by the proposed graphene-based switches as a function of the length ($\ell_{in}$) and chemical potential ($\mu_{c_{in}}$) of their central waveguide section at the fixed frequency of $28$~THz. The length of the devices ($L=1.75~\mu$m) is kept constant in all cases.(a) Graphene-based $2$D sheet switch, see Fig.~\ref{fig:_Sheet_switch}. (b) Graphene-based strip switch with $W=0.2~\mu$m, see Fig.~\ref{fig:_Strip_switch}.}
\label{fig:_L_S21_FS}
\end{figure}

In order to further study the isolation performance of the proposed switches, \figref{fig:_L_S21_FS} reports a parametric study of $S_{21}$ as a function of the length and chemical potential of the central waveguide section, at the fixed frequency of $28$~THz. The length of the devices ($L=1.75~\mu$m) is kept constant in all cases. Note that the $S_{21}$  parameter allows to evaluate the overall performance of the switches, implicitly providing information about the losses and power reflected back to the input port [see Eqs.~(\ref{Power_relationT})-(\ref{Power_relationD})]. The results have been obtained using both the TL approach (dashed line) and the full-wave solver HFSS (solid line), and are shown in \figref{fig:_L_S21_FS}(a) and \figref{fig:_L_S21_FS}(b) for ideal $2$D sheet and strip-based graphene switches, respectively. The length of the inner (switch) waveguide section $\ell_{in}$ is swept from $20$ to $500$ nm, avoiding values below $20$~nm where quantum effects may be non-negligible \cite{Christensen12}. A standing wave appears within the structure and its interference pattern varies versus $\ell_{in}$ thus explaining the oscillatory behavior of the $S21$ parameter observed in the figures. As expected, the isolation levels increase with $\ell_{in}$. As explained in \secref{Sec:Operation_principle}, when the chemical potential of the inner graphene waveguide is different from the one of the outer sections of the device, the energy propagating along the structure finds a discontinuity due to the different impedances and propagation characteristics of the plasmon modes supported by each region.  Specifically, decreasing the chemical potential of the central graphene waveguide from $0.5$~eV to $0.175$~eV leads to switches with low isolation levels. This is because the plasmon impedance is ``weakly '' affected by the chemical potential in this range (see \figref{fig:_Propagation_constant}). However, decreasing $\mu_c$ to lower values such as $0.125$~eV or $0.1$~eV leads to high isolation levels for almost any length of the inner waveguide section. Besides, it is observed that the isolation level converges when the chemical potential is further decreased. This behavior suggests that in the OFF state isolation is not governed by the fundamental plasmonic mode, but by higher order evanescent modes excited at the discontinuities between the different waveguide sections. Consequently, the transmission line approach -which by definition only consider the fundamental mode- is not able to provide accurate results in this state. However, these effects are obviously accounted for by full-wave simulations Thus, in the case of high isolation, the TL modes slightly overestimate the performance of the switch. In general, this parametric study demonstrates the excellent capabilities of graphene as a material for developing plasmonic-based switches, allowing isolation levels better than $40$~dB using graphene sections of about $500$~nm.

\begin{figure} \centering
\subfloat[]{\label{fig:_S_param_TL_fs}
\includegraphics[width=0.5\columnwidth]{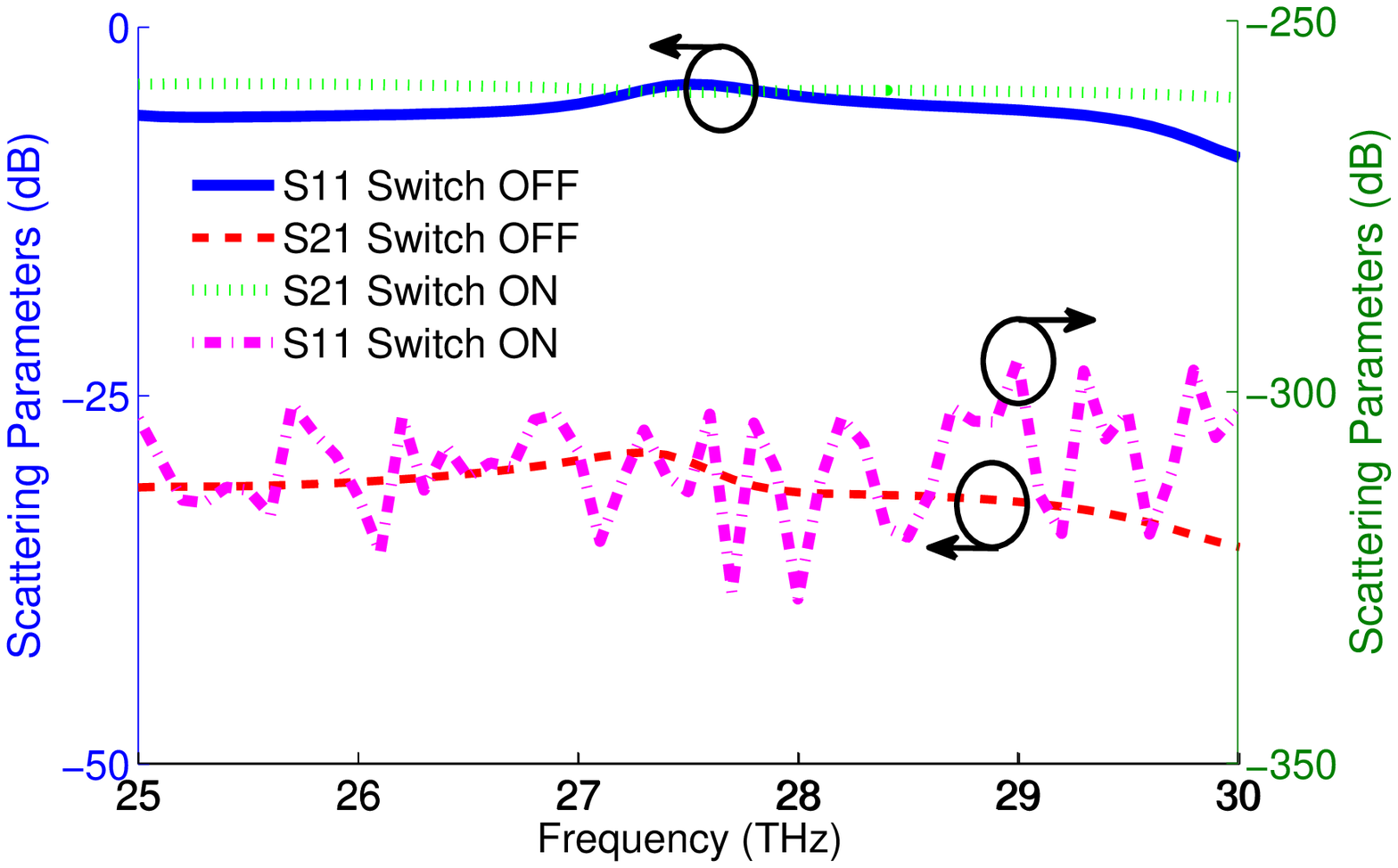}}
\subfloat[]{\label{fig:_S_param_TL_diel}
\includegraphics[width=0.5\columnwidth]{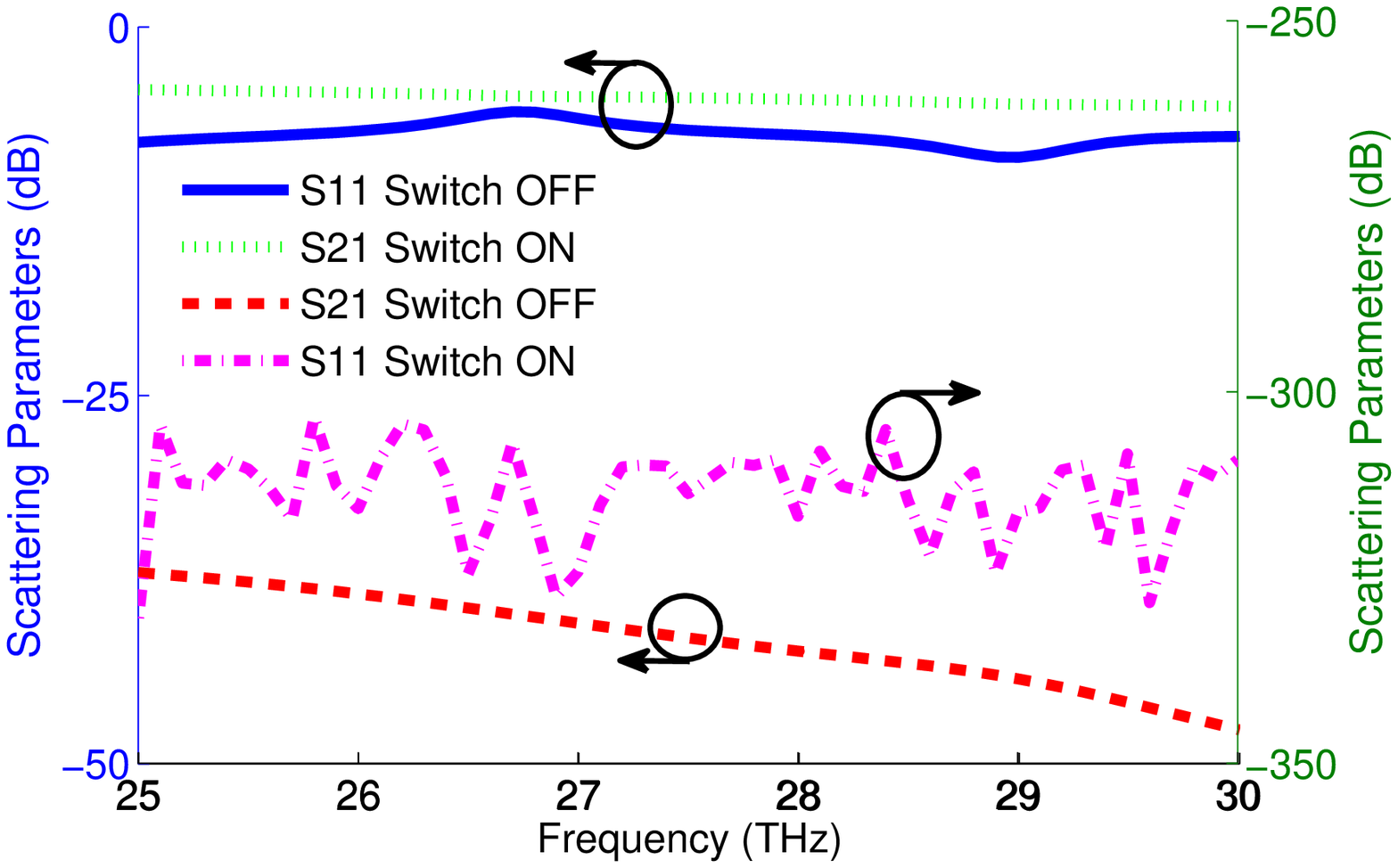}}
\caption{Simulated scattering parameters of the proposed graphene-based switches at their states ON and OFF. The parameters of the structure are $\varepsilon_r=4.0$, $L=0.7~\mu$m and $\ell_{in}=0.2~\mu$m. (a) Graphene-based $2$D sheet switch, see Fig.~\ref{fig:_Sheet_switch}. (b) Graphene-based strip switch with $W=0.2~\mu$m, see Fig.~\ref{fig:_Strip_switch}.}\label{fig:_S_param_dielectric}
\end{figure}

\begin{figure} \centering
\subfloat[]{\label{fig:_L_S21_Sheet_Diel}
\includegraphics[width=0.5\columnwidth]{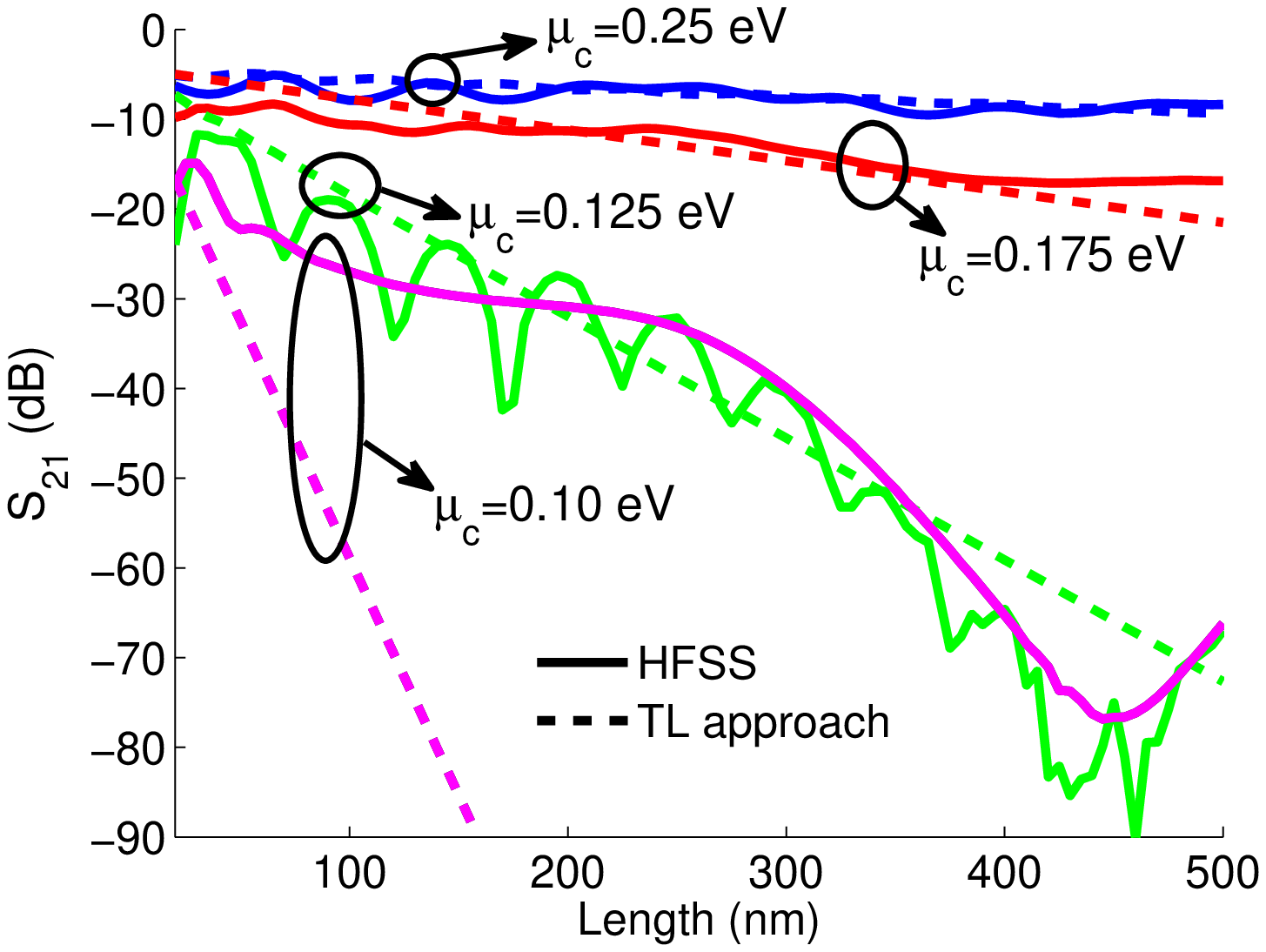}}
\subfloat[]{\label{fig:_L_S21_TL_Diel}
\includegraphics[width=0.5\columnwidth]{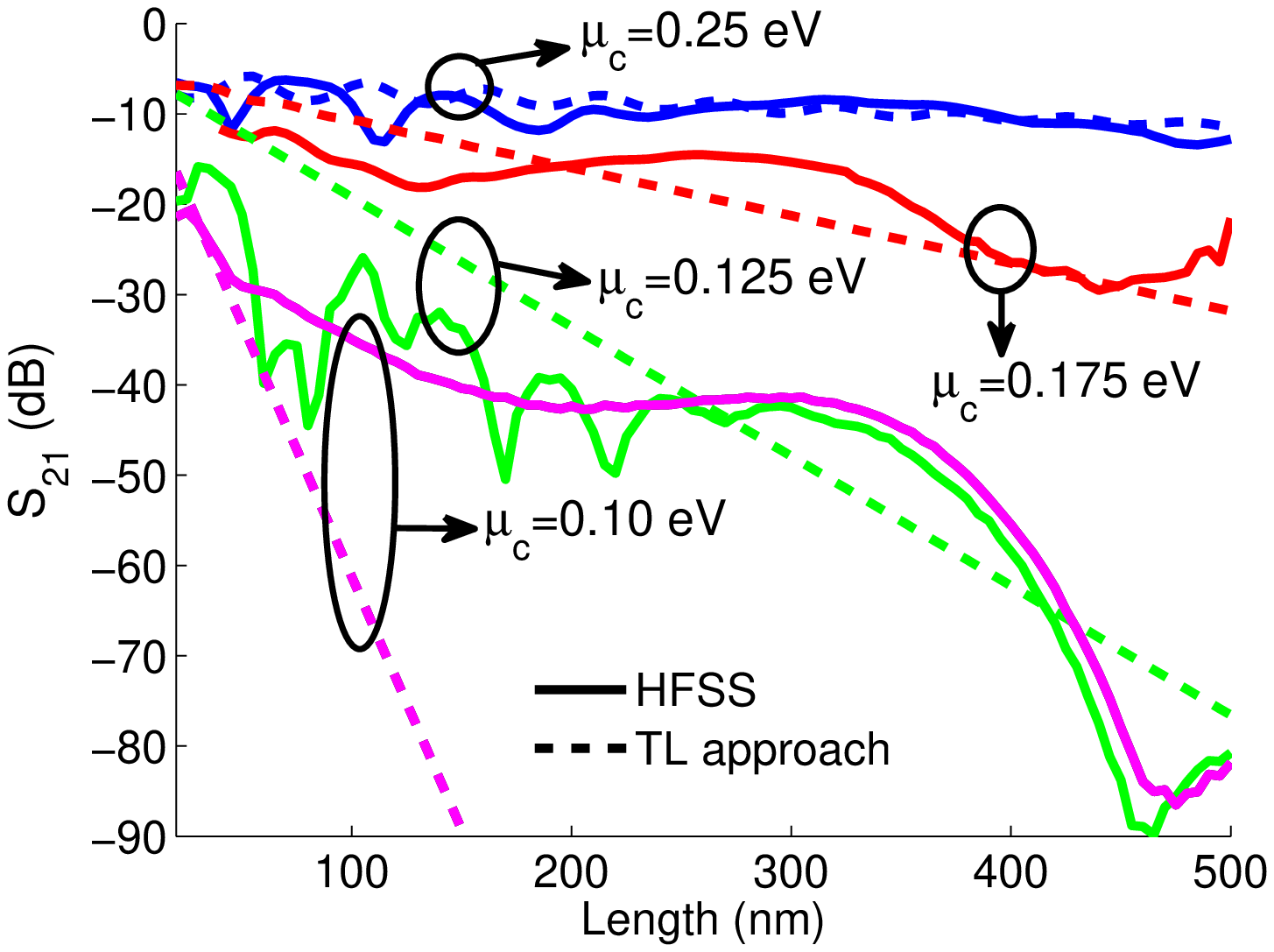}}
\caption{Parametric study of the isolation ($S_{21}$) provided by the proposed graphene-based switches as a function of the length ($\ell_{in}$) and chemical potential ($\mu_{c_{in}}$) of their central waveguide section at the fixed frequency of $28$~THz. The length of the devices ($L=1.75~\mu$m) is kept constant in all cases. The dielectric permittivity is set to $\varepsilon_r=4.0$. (a) Graphene-based $2$D sheet switch, see Fig.~\ref{fig:_Sheet_switch}. (b) Graphene-based strip switch with $W=0.2~\mu$m, see Fig.~\ref{fig:_Strip_switch}.}\label{fig:_L_S21_Dielectric}
\end{figure}

The last part of our study deals with realistic graphene-based switches taking into account the dielectric surrounding media. To this purpose, we consider graphene transferred onto a parylene of thickness $t=90$~nm and dielectric constant $\varepsilon_{r}=4.0$ at the frequencies of interest. The three polysilicon gating pads beneath the host waveguide have a thickness of $30$~nm, and the permittivity of the supporting quartz substrate is $\varepsilon_{r}=3.9$. Note that the presence of the dielectric increases the localization of the supported plasmon modes, i.e. decreases their propagation length and increases their mode confinement. The lengths of the different waveguide sections are redesigned to preserve similar insertion losses of the complete structure as in the previous example. Specifically, values of $L=0.7~\mu$m and $\ell_{in}=0.2~\mu$m are employed. Figures \ref{fig:_S_param_dielectric}(a)-\ref{fig:_S_param_dielectric}(b) report the scattering parameters of the proposed graphene-based $2$D sheet and strip switches, respectively. The results show similar insertion losses in the ON state (around $4$ dB) as compared with the free-space suspended devices of \figref{fig:_S_param_sheet}. In the OFF state, the isolation levels are around $30$ and $40$~dB for the $2$D sheet and strip switches. Note that these values are around $10$~dB lower than in the previous example, which is mainly attributed to the shorter length of the central waveguide section. In addition, \figref{fig:_L_S21_Dielectric} presents a study of the switches isolation at $28$~THz as a function of $\ell_{in}$ and $\mu_c$. Results demonstrate that the influence of these parameters on the switches performance is similar as in the case of \figref{fig:_L_S21_FS}, where the switches are suspended in free-space. Importantly, the presence of the dielectric allow to increase the isolation levels of the switches. For instance, isolation levels larger than $80$~dB are obtained for a graphene-strip based switch with a central waveguide section of $\ell_{in}=500$~nm. This is because the dielectric allows the propagation of extremely localized plasmons on graphene, whose characteristics present a wider tunable range than those of SPPs propagating on graphene transferred on a dielectric with lower permittivity. Consequently, enhanced isolation levels (or reduced device dimensions) can be obtained by designing graphene switches on dielectrics with high permittivities.

\section{Conclusions}
We have proposed and designed series switches able to dynamically control the propagation of plasmons on graphene surfaces at near infrared frequencies. Several configurations, based on $2$D graphene surfaces and strips, have been analyzed and their performance have been evaluating versus different parameters of the structures. Two different techniques, namely a transmission line approach and full-wave simulations, have been employed to characterize the switches. It has been shown that the former method provides fast results and physical insight into the problem, but it lacks of accuracy to characterize the OFF state of the devices.

Our results have demonstrated that controlling the properties of very reduced graphene areas provides extremely large isolation levels between the input and output ports. For example, isolation levels larger that $80$~dB have been achieved by using a graphene strip of just $500$~nm. In addition, it has been shown that increasing the permittivity of the surrounding media allows to increase the isolation level of the switches. These interesting features can be used to further develop guided graphene-based devices at near infrared frequencies, leading to functionalities similar to current nanophotonic plasmonic-based devices at optics.

\section*{\label{Acknoled} Acknowledgments }
This work was supported by by the Swiss National Science Foundation (SNSF) under grant $133583$ and by the EU FP$7$ Marie-Curie IEF grant ``Marconi", with ref. $300966$.
\end{document}